\documentclass[superscriptaddress,nofootinbib,amsmath,amssymb,aps]{revtex4}
\pdfoutput=1
\usepackage{braket}             
\usepackage{graphicx}           
\usepackage{bm}             
\usepackage[usenames, dvipsnames]{color}
\usepackage{hyperref}           
\usepackage{mathtools}
\usepackage[normalem]{ulem} 
\usepackage[caption=false]{subfig}
\usepackage{float}
\usepackage{graphicx}
\usepackage{soul}              
\usepackage{bigstrut,multirow,rotating}
\DeclareMathOperator{\tr}{tr}           

\DeclareMathOperator\arcsinh{arcsinh}    

\renewcommand{\Im}{\mathop{\rm Im}} 

\usepackage{colonequals} 

\usepackage[usenames, dvipsnames]{xcolor}

\definecolor{capri}{rgb}{0.0, 0.75, 1.0}

\begin{document}

\title{Entanglement harvesting in the presence of a reflecting boundary}

\author{Zhihong Liu$^{1}$, Jialin Zhang~\footnote{Corresponding author at jialinzhang@hunnu.edu.cn}$^{1}$ and Hongwei Yu~\footnote{Corresponding author at hwyu@hunnu.edu.cn}}
\affiliation{Department of Physics and Synergetic Innovation Center for Quantum Effects and Applications, Hunan Normal University, Changsha, Hunan 410081, China}
\date{\today}
\begin{abstract}
We study, in the framework of  the entanglement harvesting protocol,
the entanglement harvesting of both a pair of inertial and uniformly accelerated
detectors locally interacting with  vacuum massless scalar fields subjected
to a perfectly  reflecting plane boundary. We find that the presence of the boundary generally degrades  the harvested entanglement when two detectors are
very close to the boundary. However, when the distance between
detectors and the boundary becomes comparable to the interaction duration parameter,  the amount of the harvested entanglement  approaches a
peak, which even goes beyond that  without a boundary.  Remarkably,  the parameter space of the detectors' separation and the magnitude of acceleration that allows
 entanglement harvesting to occur is enlarged  due to the presence of the boundary. In this sense,  the boundary plays a double-edged role on entanglement harvesting,
 degrading in general the harvested entanglement  while  enlarging the entanglement harvesting-achievable parameter space.
 A comparison of three different acceleration scenarios of the detectors with respect to the boundary, i.e.,  parallel, anti-parallel and mutually
perpendicular acceleration, shows that the phenomenon of entanglement harvesting crucially depends on the acceleration,  the separation between two detectors and the detectors' distance from the boundary.

 \end{abstract}

\maketitle


\section{Introduction}
Quantum entanglement has
been considered as  a crucial physical resource in quantum
information science  such as  quantum
communication~\cite{RC,Bh}, quantum teleportation~\cite{Bg}, quantum
cryptography~\cite{Ea} and dense coding~\cite{Bh}, etc,  and it has been extensively studied in various physical aspects. Recently, a lot
of  interest has been attracted in the role played by
entanglement in a variety of physical contexts, such as the critical
phenomena in condensed matter
systems~\cite{Osterloh:2002,Vidal:2003,Amico:2008}, the description
of non-classical states of light~\cite{LM:1995,Weedbrook:2012}, the
explanation for the origin of black hole
entropy~\cite{Bombelli:1986,Calla:1994,Srednicki:1993} and the anti-de
Sitter/conformal field theory correspondence~\cite{Ryu:2006}.

It has been  realized that  vacuum can be a resource of entanglement for the vacuum state of a free quantum field can maximally violate Bell's inequalities as was shown in the formal algebraic quantum field theory~\cite{Summers:1985,Reznik:2005}, and a pair of initially uncorrelated detectors can
extract  entanglement from  vacuum via locally interacting
with vacuum fields~\cite{VAN:1991,Reznik:2003}. This  phenomenon of entanglement extraction has been extensively studied
in  an operational approach  employing the Unruh-DeWitt
(UDW) detector
model~\cite{Steeg:2009,Olson:2011,BLHu:2012,EDU:2012-2,Martin-Martinez:2014gra,EDU:2013,Nambu:2013,Pozas-Kerstjens:2015,EDU:2016-1,EDU:2016-2,Zhjl:2018,Zhjl:2019,Ng:2018,Ng:2018-2,Zhjl:2020},
which is now known as the entanglement harvesting protocol~\cite{Salton-Man:2015}. A lot of studies  have demonstrated  that the phenomenon
of entanglement harvesting  involves  a combination
of relativistic and quantum effects, and it is  sensitive to
the spacetime
topology~\cite{Steeg:2009,EDU:2016-1,Zhjl:2018,Zhjl:2019,Ng:2018-2},
intricate
motions of detectors~\cite{Zhjl:2020}, and  even cosmology~\cite{EDU:2012-2,Martin-Martinez:2014gra}.  More recently,
the entanglement  harvesting in some
special circumstances has been investigated, such as  in the
background with black holes~\cite{Zhjl:2018}, in the presence of gravitational
waves~\cite{XQ:2020} and  near the horizon mimicked by imposing a moving
mirror as the boundary~\cite{CW:2019,CW:2020}. In particular,  the entanglement harvesting  has been studied for inertial detectors in 1+1 dimensional  moving  mirror spacetimes in Ref.~\cite{CW:2019}, where it was argued that  there exists an entanglement inhibition phenomenon similar to that found for black holes and the concrete  harvesting process is sensitive to the mirror trajectory. While, for an accelerated  mirror moving along a particular trajectory parameterized  by the product-log function,  the mimicked effects of horizon  on entanglement harvesting  have been discussed in Ref.~\cite{CW:2020}, which reveals a sensitivity of the entanglement harvested to the dynamics of the trajectories and an insensitivity of entanglement to the sign of radiation flux  emitted  by such an accelerated mirror.

In this paper, we plan to investigate the entanglement harvesting for two UDW detectors at rest or uniformly accelerated near a static perfectly reflecting boundary rather than a   drifting mirror that is used to mimic  a  dynamical spacetime. This would allow  us to consider how the reflecting boundary and the  motion status of detectors affect the entanglement harvesting phenomenon in a more realistic 1+3 dimensional spacetime. The phenomenon of  entanglement  harvesting in the circumstance with accelerated detectors  necessarily involves both the relativistic effects due to the non-inertial motion of the detectors and quantum effects arising from the modification of  quantum vacuum fluctuations of the scalar fields by the presence of a boundary which the detectors are coupled to.
Let us note here that the most fascinating effect associated with the uniform acceleration is the Unruh effect, which attests that accelerated detectors in vacuum will observe a thermal radiation spectrum of particles. The thermal noise due to the Unruh effect is generally expected to drive the accelerated
detectors to decohere. Therefore,  the behavior of entanglement
harvesting therein would be an interesting topic, since it is expected to be affected by both the presence of the boundary and the accelerated motion of the detectors which is connected the Unruh effect.
Indeed, on one hand, a lot of studies have revealed the influence of
acceleration on quantum entanglement via various detector models,
including  two-level detectors or UDW
models~\cite{Reznik:2003,Salton-Man:2015}, harmonic
oscillators~\cite{BLH:2008,BLH:2010}, and wave packets~\cite{Ahmadi:2016,Crochowski:2019}.  In particular,
it was
argued in Ref.~\cite{Salton-Man:2015} that the entanglement extraction can be enhanced for two UDW detectors in anti-parallel acceleration in comparison with those
in  inertial motion or parallel acceleration. On the other
hand, it has been demonstrated that the presence of boundaries in
flat spacetime, which changes the spacetime topology and
causes modification of fluctuations of quantum
fields~\cite{Birrell:1984}, brings new features to novel quantum effects
associated acceleration, such as the Casimir-Polder
 interaction~\cite{Rizzuto:2007,Passante:2007,Zhu:2010}, the modified
radiative properties of accelerated
atoms~\cite{Yu:2005ad,Yu:2006kp,Rizzuto:2009}, the geometric phase~\cite{Zhjl:2016}
and the modified entanglement dynamics~\cite{Zhjl:2007,Cheng:2018}.

Therefore,  questions arise naturally as to  what role  the
boundary would play in performing  the entanglement harvesting
protocols via  a pair of inertial as well as uniformly accelerated detectors and what would happen to the entanglement harvesting for
 two  detectors  in different acceleration scenarios near the boundary. To address these questions,  we will first consider a pair of inertial detectors and then a pair of uniformly accelerated UDW
detectors, which are initially prepared in a separable
state and  locally interact with the vacuum massless scalar fields  near a perfectly
reflecting plane boundary.  For the case of accelerated detectors,  three distinct acceleration scenarios, i.e, parallel acceleration and anti-parallel acceleration with respect to the boundary as well as mutually perpendicular acceleration in a plane parallel to the boundary, will be examined.
According to the entanglement harvesting
protocol, the  reduced density of two detectors
can be  written in an  $X$-type form, and the concurrence as a
measure of entanglement, which is employed to characterize the
amount of the entanglement harvested from vacuum by such
detectors, can be calculated  for a chosen  switching
function.

The paper is organized as follows. In the next section,  basic
formulae for the UDW detectors locally interacting with vacuum
scalar fields are reviewed.  In Sec.III,
we mainly study the influence of a boundary  on the transition
probabilities of uniformly accelerated detectors interacting with
the quantum fields in a finite duration. In Sec. IV, we  consider
the entanglement harvesting for both two inertial detectors and two detectors
moving along  different acceleration trajectories near the boundary,
including those of parallel acceleration, anti-parallel
acceleration and acceleration in mutually perpendicular directions  in a plane parallel to the boundary.
The influence of the boundary on the behavior of  the entanglement harvesting
is examined in detail, with a cross-comparison of the results obtained. Finally,
we end up with conclusions in Sec.V.

Throughout this paper we adopt the natural units in which $\hbar=c=1$  for
convenience.
\section{The basic formulas}
Without loss of generality, the two-level point-like detector is
treated as the UDW module with a ground state $|0_{D}\rangle$ and
excited state $|1_{D}\rangle$, which locally  interacts with the
massless scalar field $\phi(x_D)$. Here, $x_D$ denotes the
coordinates of spacetime with the subscript $D$ specifying which UDW
detector we are considering.  Let us suppose that the classical spacetime
trajectory of the detector is parameterized by its proper time
$\tau$. Then the  interaction Hamiltonian for such a detector in the
interaction picture takes the following form
 \begin{equation}\label{Int1}
 H_{D}(\tau)=\lambda \chi(\tau)\left[e^{i \Omega_D \tau} \sigma^{+}+e^{-i \Omega_D\tau} \sigma^{-}\right] \phi\left[x_{D}(\tau)\right]\;,
 \end{equation}
where $\lambda$ is the coupling strength, $\Omega_D$ is the energy gap of the detector,
$\sigma^{+}=|1_{D}\rangle\langle0_{D}|$  and
$\sigma^{-}=|0_{D}\rangle\langle1_{D}|$ denote the ladder operators
of the SU(2) algebra, and $\chi(\tau)=\exp [-{\tau^{2}}/(2
\sigma_D^{2})]$ is the Gaussian switching function  which controls the
duration of interaction via  parameter $\sigma_D$.

Before the interaction begins, we assume that two UDW  detectors (labeled
$A$ and $B$) are in their ground state and the field  in a
vacuum state $|0_M\rangle$. Then the joint state of the detectors
and the field can be written as
$\ket{\Psi}=\ket{0_A}\ket{0_B}\ket{0_M}$. According to the
detector-field interaction Hamiltonian~(\ref{Int1}), the finial
state of the system (two detectors plus the field) is
given by
\begin{equation}\label{psi-f}
\ket{\Psi_f}:={\cal{T}} \exp\Big[-i\int{dt}\Big(\frac{d\tau_A}{dt}H_A(\tau_A)+
\frac{d\tau_B}{dt}{H_B}(\tau_B)\Big)\Big]\ket{\Psi}\;,
\end{equation}
where ${\cal{T}}$ denotes the time ordering operator.  For
simplicity, the two detectors  are assumed to be completely identical with a fixed but not too large energy gap,
i.e., $\Omega=\Omega_A=\Omega_B$ and $\sigma=\sigma_A=\sigma_B$, in the following discussions.
Based on the perturbation theory,  the density matrix for the finial state of the two
detectors can be obtained from Eq.~(\ref{psi-f}) by tracing out the
field degrees of freedom, and after some  algebraic manipulations, it
takes the following form  in the basis
$\{\ket{0_A}\ket{0_B},\ket{0_A}\ket{1_B},\ket{1_A}\ket{0_B},\ket{1_A}\ket{1_B}\}$
~\cite{EDU:2016-1,Zhjl:2018,Zhjl:2019}
\begin{align}\label{rhoAB}
\rho_{AB}:&=\tr_{\phi}\big(\ket{\Psi_f}\bra{\Psi_f}\big)\nonumber\\
&=\begin{pmatrix}
1-P_A-P_B & 0 & 0 & X \\
0 & P_B & C & 0 \\
0 & C^* & P_A & 0 \\
X^* & 0 & 0 & 0 \\
\end{pmatrix}+{\mathcal{O}}(\lambda^4)\;,
\end{align}
where the transition probability $ P_D$ reads
\begin{equation}\label{PAPB}
 P_D:=\lambda^{2}\iint d\tau d\tau' \chi(\tau) \chi(\tau') e^{-i \Omega(\tau-\tau')}
 W\left(x_D(t), x_D(t')\right)\quad\quad D\in\{A, B\}\;,
\end{equation}
and quantities $C$ and $X$ which characterize nonlocal correlations  are given by
\begin{align}
C &:=\lambda^{2} \iint d \tau d \tau^{\prime} \chi(\tau) \chi(\tau')
e^{-i \Omega(\tau-\tau')} W\left(x_{A}(t), x_{B}(t')\right)\;,
\end{align}
\begin{align}\label{xxdef}
X:=-\lambda^{2} \iint d\tau d \tau' \chi(\tau)\chi(\tau')  e^{-i\Omega( \tau+\tau')}
\Big[\theta(t'-t)W\left(x_A(t),x_B(t')\right)+\theta(t-t')W\left(x_B(t'),x_A(t)\right)\Big]\;,
\end{align}
where  $W(x,x'):=\bra{0_M}\phi(x)\phi(x')\ket{0_M}$ is the
Wightman function of the field and $\theta(t)$
represents the Heaviside theta function. Note that the detector's
coordinate time is a function of its proper time,
i.e., $t=t(\tau)$ and $t'=t'(\tau')$, in the above equations.

According to the entanglement harvesting protocol, we can employ the concurrence as a measure of entanglement~\cite{WW}, which
specifically quantifies the entanglement harvested by the detectors
via local  interaction with the fields. For an $X$-type density
matrix~(\ref{rhoAB}),  the concurrence takes a
simple form~\cite{EDU:2016-1,Zhjl:2018,Zhjl:2019}
\begin{equation}\label{condf}
\mathcal{C}(\rho_{A B})=2 \max \Big[0,|X|-\sqrt{P_{A}
P_{B}}\Big]+\mathcal{O}(\lambda^{4})\;.
\end{equation}
The concurrence $\mathcal{C}(\rho_{A B})$ is  dependent only on the
nonlocal correlation X and the transition probabilities. As a result, the
Wightman function of the scalar fields  plays a crucial role in
 entanglement harvesting.
In what follows, We will examine  the entanglement harvesting
phenomenon for both a pair  of  inertial and uniformly accelerated detectors near a
perfectly reflecting plane boundary, focusing on the influence of
the presence of the boundary.

\section{The transition probabilities of detectors near the reflecting boundary}

Let us first analyze the transition probabilities. To facilitate the discussion, we assume that a  plane boundary is
located  at $z = 0$, and the uniformly
accelerated UDW detector  is moving along the trajectories with a
distance $\Delta{z}$  away from the boundary, that is
\begin{equation}\label{a-trj}
x_D:=\{t={a^{-1}}\sinh(a\tau),\;x={a^{-1}}\cosh (a \tau),\;
 y=0,\; z=\Delta{z}\}\;,
\end{equation}
where $a$ is the  proper acceleration and $\tau$ is the detector's
proper time.
The Wightman function for vacuum massless scalar
fields in four dimensional Minkowski spacetime in the presence of a reflecting
boundary is, according to the method of images, given by~\cite{Birrell:1984}
\begin{align}\label{wigh-1}
W\left(x_D, x'_D\right)=&-\frac{1}{4 \pi^{2}}\Big[\frac{1}{(t-t'-i
\epsilon)^{2}-(x-x')^{2}-(y-y')^{2}
-(z-z')^{2}}\nonumber\\
&-\frac{1}{(t-t'-i \epsilon)^{2}-(x-x')^{2}-(y-y')^{2}
-(z+z')^{2}}\Big]\;.
\end{align}
Substituting  trajectory~(\ref{a-trj}) and  Eq.~(\ref{wigh-1}) into
Eq.~(\ref{PAPB}), we have, after some manipulations (see
Appendix)
\begin{align}\label{PD-1}
P_{D} =&\frac{\lambda^{2} a \sigma}{4 \pi^{3 / 2}} \int_{0}^{\infty}
d\tilde{s} \frac{\cos (\tilde{s} \beta)e^{-\tilde{s}^{2}
\alpha}\left(\sinh ^{2}\tilde{s}-\tilde{s}^{2}\right)}{\tilde{s}^{2}
\sinh^{2} \tilde{s}}+\frac{\lambda^{2} a \sigma}{4
\pi^{{3}/{2}}}{\rm{PV}} \int_{0}^{\infty} d \tilde{s}
\frac{\cos(\beta{\tilde{s}}) e^{-\tilde{s}^{2} \alpha}}{\sinh
^{2}\tilde{s}-a^{2} \Delta z^{2}}\nonumber
\\&+\frac{\lambda^{2}}{4 \pi}\left[e^{-\Omega^{2}
\sigma^{2}}-\sqrt{\pi} \Omega \sigma \operatorname{Erfc}(\Omega
\sigma)\right]+\frac{\lambda^{2} a \sigma}{4
\sqrt{\pi}}\frac{e^{-\alpha{\tilde{s}^2}}\sin(\beta{\tilde{s}})}{\sinh(2\tilde{s})}\Bigg|_{\tilde{s}=\arcsinh(a\Delta{z})}\;,
\end{align}
where  $\beta=2\Omega/ a$, $\alpha=1 /(a \sigma)^{2}$ and $\rm{Erfc}(x):=1-\rm{Erf}(x)$ with the error function $\rm{Erf}(x):=\int_0^x2e^{-t^2}dt/\sqrt{\pi}$.

An analytical  result of Eq.~(\ref{PD-1}) can  hardly be
found. Fortunately,  numerical evaluations can be implemented for a finite duration of interaction
( finite nonzero $\sigma$ ). It is worth noting that the second and
fourth term in Eq.~(\ref{PD-1}) arise from the image part of the
Wightman function, which are dependent on the distance between the boundary and the detector  $\Delta{z}$,  while the remaining  two terms are just those of the
transition probabilities of uniformly accelerated detectors in free
Minkowski spacetime~\cite{Zhjl:2020}. In the limit of
$a\rightarrow0$, the transition probabilities become
\begin{align}\label{PDa0}
P_{D} &=\frac{\lambda^{2}}{4 \pi}\left[e^{-\Omega^{2} \sigma^{2}}-\sqrt{\pi}
\Omega \sigma \operatorname{Erfc}(\Omega
\sigma)\right]+\frac{\lambda^2\sigma}{4\pi^{3/2}}\int_{-\infty}^{\infty}{ds}\frac{e^{-i\Omega{s}}e^{-s^2/(4\sigma^2)}}{(s-i\epsilon)^2-4\Delta{z}^2}\nonumber\\
&=\frac{\lambda^{2}}{4 \pi}\left[e^{-\Omega^{2} \sigma^{2}}-\sqrt{\pi}
\Omega \sigma \operatorname{Erfc}(\Omega
\sigma)\right]-\frac{\lambda^{2}\sigma{e}^{-\Delta{z}^2/\sigma^2}}{8\sqrt{ \pi}\Delta{z}}\left\{\Im\Big[e^{2i\Omega\Delta{z}}\rm{Erf}\Big(i\frac{\Delta{z}}{\sigma}+\Omega\sigma\Big)\Big]-\sin(2\Omega\Delta{z})\right\}\;.
\end{align}

\begin{figure*}[!ht]
\subfloat[]{\label{PA11}
\includegraphics[width=0.51\textwidth]{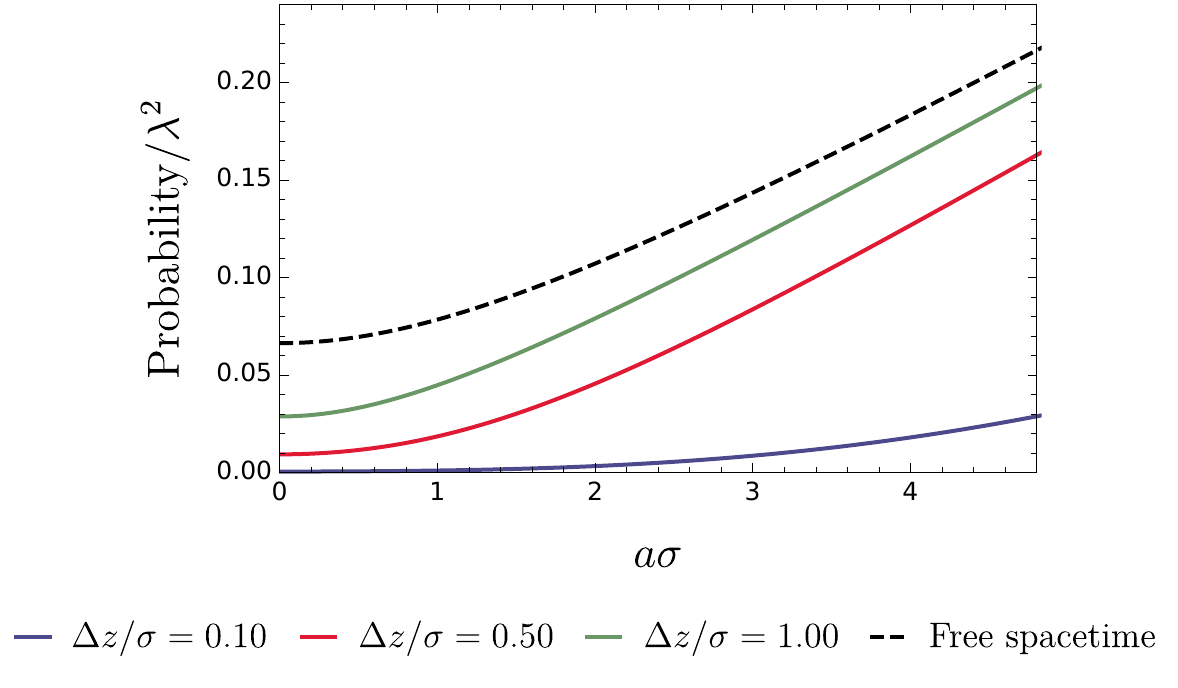}}~\subfloat[]{\label{PA22}
\includegraphics[width=0.40\textwidth]{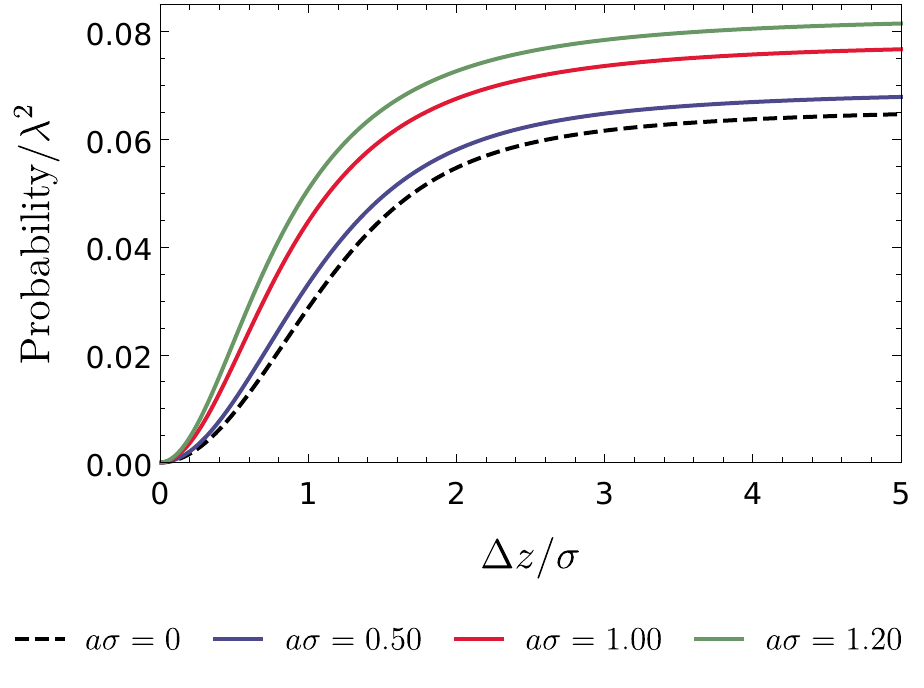}}
\caption{The transition probability of UDW detectors is plotted as a
function of the  acceleration  with parameters $\Omega\sigma=0.10$
and $\Delta z / \sigma=\{0.10,0.50,1.00\}$ in (a), and plotted as a
function of $\Delta{z} /\sigma$ with $\Omega\sigma=0.10$ and
$a\sigma=\{0,0.50,1.00,1.20\}$ in (b).  For convenience, all other
physical parameters are expressed  in the unit of the
interaction duration parameter $\sigma$.  Note that the dashed lines indicate the
case of without any boundary in (a) and  inertial detectors at rest in (b), respectively.}\label{PA}
\end{figure*}

We show our numerical evaluation of  Eq.(\ref{PD-1}\&\ref{PDa0}) in  Fig.~(\ref{PA}).
As we can see there,  the transition probabilities for a
finite duration generally increase as the acceleration
increases, which is consistent with what we expect that
high acceleration causes strong thermalization. However, the
thermalization is suppressed by the presence of the boundary. Especially, when the detector is very close to the boundary ($\Delta{z}/\sigma\ll1$),
the transition probabilities  become very small.
   At the same time, one can also observe that the transition
probabilities seem to be an increasing function of $\Delta{z}/\sigma$
which flatten up when $\Delta{z}/\sigma$ becomes relatively large.

\section{  Entanglement harvesting  near the reflecting boundary}
In this section, we will examine the phenomenon of the entanglement
harvesting, paying particular attention to
 the influence of the boundary. We start with a discussion of how the reflecting boundary affects the entanglement harvested by two inertial detectors at rest at a certain fixed distance $\Delta{z}$ from the boundary, followed by an analysis of two uniformly accelerated detectors.
For convenience of comparing  the boundary influence in these two circumstances,  the trajectories of two accelerated detectors  are assumed  to align parallel to the boundary plane with a  distance $\Delta{z}$ away from it as well. The effects of the boundary on the entanglement harvesting in  three different acceleration scenarios, i.e.,  parallel acceleration, anti-parallel acceleration and  mutually perpendicular acceleration, are to be analyzed  in detail.

\subsection{Entanglement harvesting for  inertial detectors }
\begin{figure}[!htbp]
\centering
\subfloat[]{\label{cona011}\includegraphics[width=0.40\linewidth]{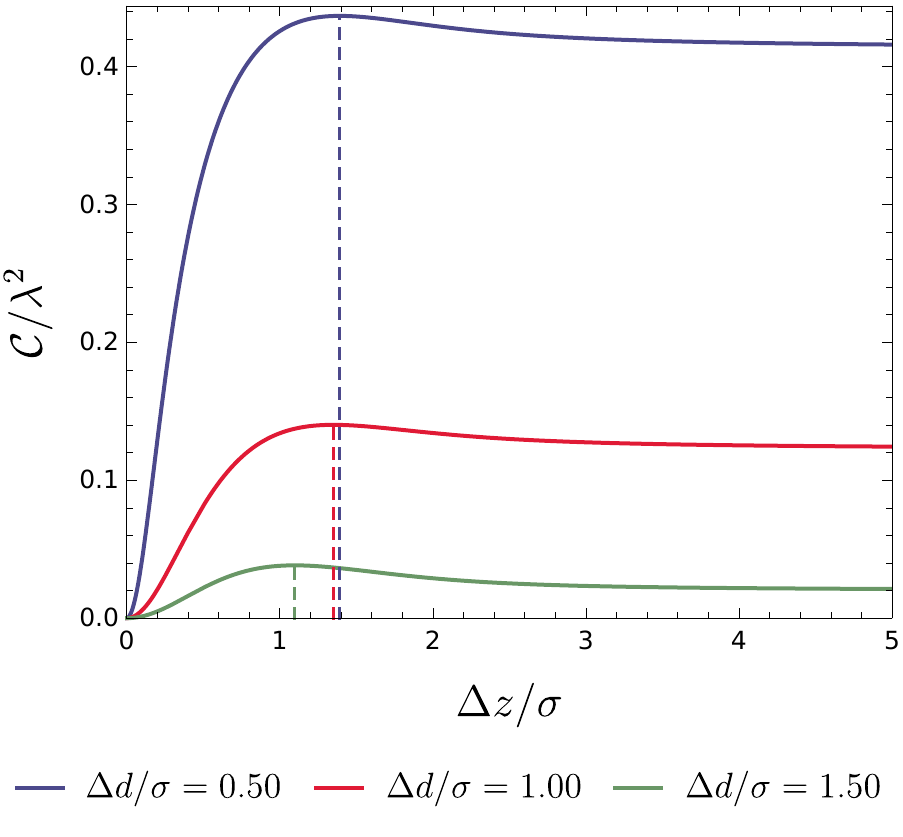}}\;\;
 \subfloat[]{\label{cona022}\includegraphics[width=0.40\linewidth]{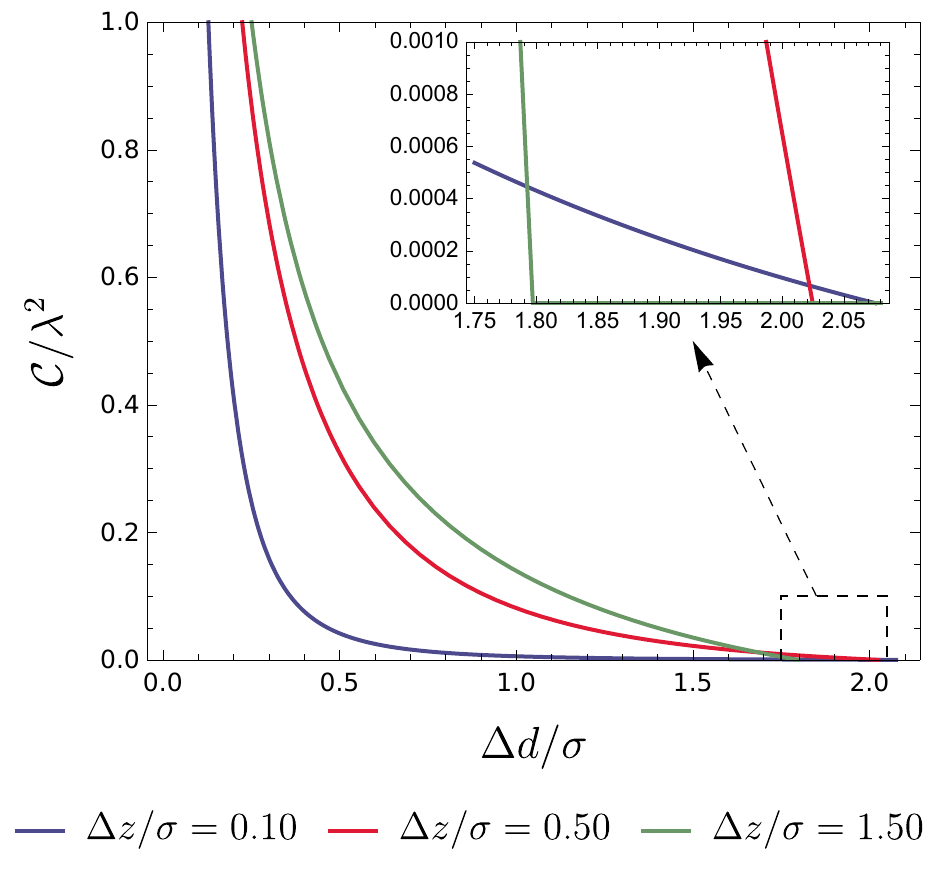}}
\caption{The entanglement harvested by   detectors at rest  in the presence of a boundary. (a) The concurrence as a function of
  $\Delta{z}/\sigma$ is plotted for  $\Delta{d}/\sigma=\{0.50,1.00,1.50\}$. The dashed line indicates  the  peak of the concurrence. (b) The concurrence as a function of $\Delta{d}/\sigma$ is plotted  for $\Delta{z}/\sigma=\{0.10,0.50,1.50\}$. Here, we have set $\Omega\sigma=0.10$.  The  concurrence $\mathcal{C}(\rho_{A B})$  rapidly  falls off  as the separation $\Delta{d}$ increases for a fixed distance from the boundary.  In the limit of $\Delta{z}\rightarrow0$, the concurrence  approaches zero. }\label{cona0}
 \end{figure}

 For two detectors at rest which are aligned parallel  to  the  reflecting boundary with a fixed distance $\Delta{z}$ from it and separated by a distance  $\Delta{d}$, their trajectories  can,  without loss of generality,   be written as
\begin{equation}\label{rest-trj}
x_A:=\{\tau_A\;,~x=0\;,~y=0\;,~z=\Delta{z}\}\;, x_B:=\{\tau_B\;,~x=\Delta{d}\;,~y=0\;,~z=\Delta{z}\}\;.
\end{equation}
 In order to examine the effect of the boundary on the entanglement harvesting for inertial detectors, we should first compute  the transition probability $P_D$ given by Eq.~(\ref{PAPB}) and the relevant  non-local correlation term $X$ given by Eq.~(\ref{xxdef}) along the trajectories~(\ref{rest-trj}).  The analytical expression of the transition probability of a static detector has already been given by Eq.~(\ref{PDa0}), and  $X$ particularized here as $X_0$, can be obtained straightforwardly via the method of Cauchy principle value,
\begin{equation}\label{xx0}
X_0=\frac{i\lambda^2\sigma}{4\sqrt{\pi}}e^{-\frac{\Delta{d}^2+4\sigma^4\Omega^2}{4\sigma^2}}
\left\{\frac{{\rm{Erfc}}\Big[i{\sqrt{\Delta{d}^2+4\Delta{z}^2}}/{(2\sigma)}\Big]}{\sqrt{\Delta{d}^2+4\Delta{z}^2}}e^{-{\Delta{z}^2}/{\sigma^2}}
-\frac{{\rm{Erfc}}\Big[i{\Delta{d}}/{(2\sigma)}\Big]}{\Delta{d}}\right\}\;.
\end{equation}
Then the concurrence can be  found by substituting Eq.~(\ref{PDa0}) and Eq.~(\ref{xx0}) into Eq.~(\ref{condf}).  In order to show the properties of the entanglement harvesting, we plot the concurrence  as a function of the distance $\Delta{z}$ and the separation $\Delta{d}$ in Fig.~(\ref{cona0}), respectively.

 It is worth pointing out that $X_0$  will be vanishingly small when the two detectors are placed infinitely close to the reflecting boundary and so will be the concurrence, suggesting that the entanglement harvesting will be greatly  inhibited as is illustrated by Fig.~(\ref{cona011}). More interestingly, one may  find a peak of the extracted
entanglement approximately at the position where $\Delta{z}$ is
comparable to the interaction duration parameter $\sigma$ for a not
too large fixed $\Delta{d}/\sigma$ \footnote{If $\Delta{d}/\sigma$
were too large, the extracted entanglement would vanish.}. At  large
enough $\Delta{z}/\sigma$, the concurrence  asymptotically
approaches its free space value as expected.  Let us note here that
these features, e.g.,  the appearance of peaks of the extracted
entanglement,  have also been found for inertial detectors in 1+1
dimensional mirror spacetimes  in Ref.~\cite{CW:2019}.

\begin{figure}[!htbp]
  \centering
 \includegraphics[width=0.50\linewidth]{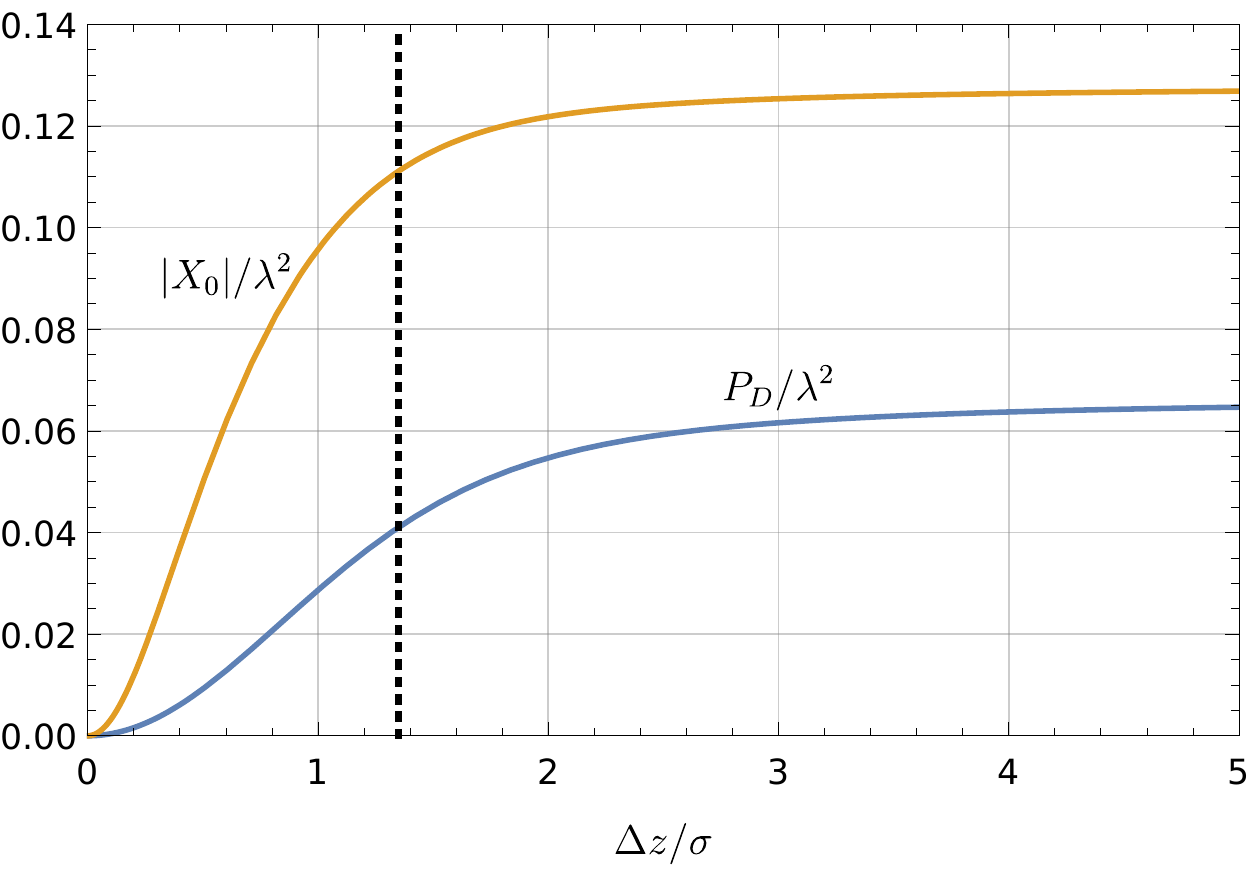}
  \caption{ The nonlocal correlation term  $|X_{0}|$   and the transition probability $P_{D}$ are plotted as a function of the distance between the rest
   detectors and the boundary, $\Delta{z}$.  The  vertical dashed line ($\Delta{z}/\sigma\approx1.352$) indicates where the maximum difference between $|X_{0}|$ and $P_{D}$ occurs. Here, we have set the energy gap $\Omega\sigma=0.10$ and  the separation between such two rest detectors $\Delta{d}/\sigma=1.00$.}\label{XX-PA-Z}
 \end{figure}

In order to gain an understanding to this
property,  we show  the behaviors  of $P_D$ and $|X_{0}|$ versus $\Delta{z}/\sigma$ in Fig.~(\ref{XX-PA-Z}). As we can see,  the  reflecting boundary would in general restrain both the transition probabilities $P_D$ and  the non-local correlation of the fields $X_0$.   However,  as the distance $\Delta{z}$ increases to approach  $\sigma$, the degree of suppression is different (see the vertical dashed line in  Fig.~(\ref{XX-PA-Z})). That is, at the beginning,  the difference between $|X_0|$ and $P_D$ grows as the distance increases. However, when the distance  grows across  $\sigma$, $|X_{0}|$  no longer increases significantly  with  increasing $\Delta{z}/\sigma$,  while the transition probability $P_D$ still significantly increases for a while.  As a result,  the difference between $|X_{0}|$ and $P_D$  becomes a little smaller, resulting in a peak in the difference therein. As, according to the definition  in Eq.~(\ref{condf}), the concurrence is in fact a competition between the nonlocal correlation $X$ and the transition probabilities, the above analysis explains why the harvested entanglement peaks when the distance is comparable to the parameter $\sigma$. Remarkably,  the peak value of the concurrence is even larger than that without a boundary, suggesting that the detectors may  harvest more entanglement than when there is no boundary.

Fig.~(\ref{cona022}) shows that the concurrence for a fixed distance between detectors and the boundary  would  rapidly degrade as the detectors' separation  increases.
This arises from the fact that $X_0$ is exponentially suppressed as  $\Delta{d}$  increases (see Eq. (\ref{xx0})), and is consistent with our intuition  that the correlation of the detectors would weaken as their inter-separation grows. Once the inter-separation increases beyond a certain value, the concurrence will virtually vanish  and entanglement harvesting essentially on longer occurs.  So,  there exists an entanglement  harvesting-achievable range for $\Delta{d}$, which, as shown in the subfigure of Fig.~(\ref{cona022}), is sensitive to the distance from the boundary.

\begin{figure}[!htbp]
  \centering
 \includegraphics[width=0.50\linewidth]{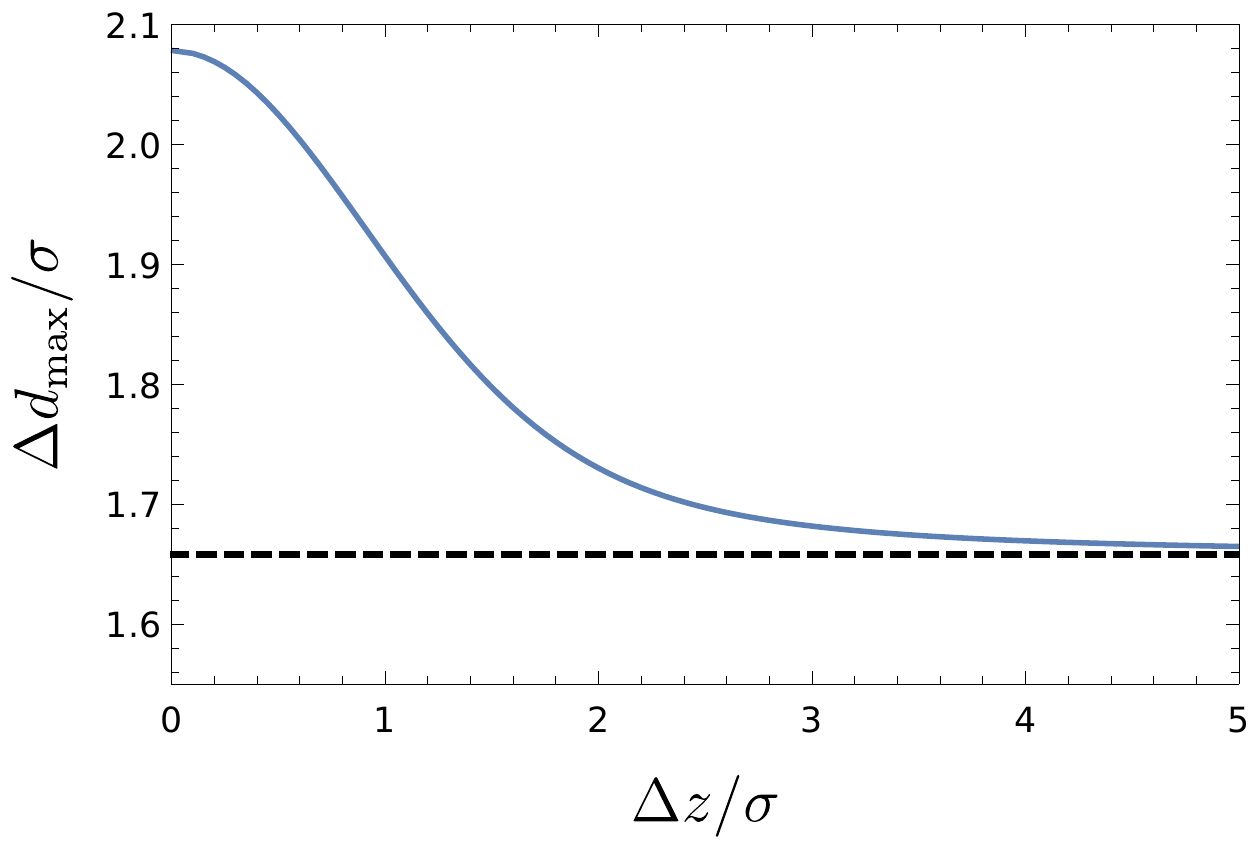}
  \caption{ The maximum harvesting-achievable separation, $\Delta{d}_{\rm{max}}$, between two detectors is plotted as
a function of the distance between detectors and the boundary with  $\Omega\sigma=0.10$. The dished line indicates the case of a free space without any boundary. }\label{dmax-a0-Z}
 \end{figure}

To better understand the influence of the presence of the boundary on the harvesting-achievable range of
the separation $\Delta{d}$ where entanglement harvesting is possible. Here, we use $\Delta{d}_{\rm{max}}$  to  stand for the  maximum value (or critical value) of the separation $\Delta{d}$, beyond which entanglement harvesting does not occur any more.  From Eq.~(\ref{PDa0}) and Eq.~(\ref{xx0}), we obtain  the plot of  $\Delta{d}_{\rm{max}}$ as a function of $\Delta{z}$ (see Fig.~(\ref{dmax-a0-Z})). One can see that $\Delta{d}_{\rm{max}}$   is obviously a  decreasing function of $\Delta{z}/\sigma$, which means that the presence of the boundary could  enlarge the harvesting-achievable range. It should be pointed out here that a $\Delta{d}_{\rm{max}}$  on the boundary plane  would make no physical sense. This is because the corresponding Wightman function Eq.~(\ref{wigh-1}) approaches zero in the limit of $\Delta{z}\rightarrow0$, and as a consequence  the concurrence then vanishes and does not depend on the parameter $\Delta{d}$.

Now we are in a position to explore the effects of the boundary on the entanglement harvesting of uniformly accelerated detectors. In particular, we will  study  three acceleration scenarios  and make a comparison with the inertial situation.

\subsection{Entanglement harvesting for uniformly accelerated detectors }
\subsubsection{ Parallel acceleration}
\begin{figure}[!htbp]
  \centering
 \subfloat[] {\includegraphics[width=0.35\linewidth]{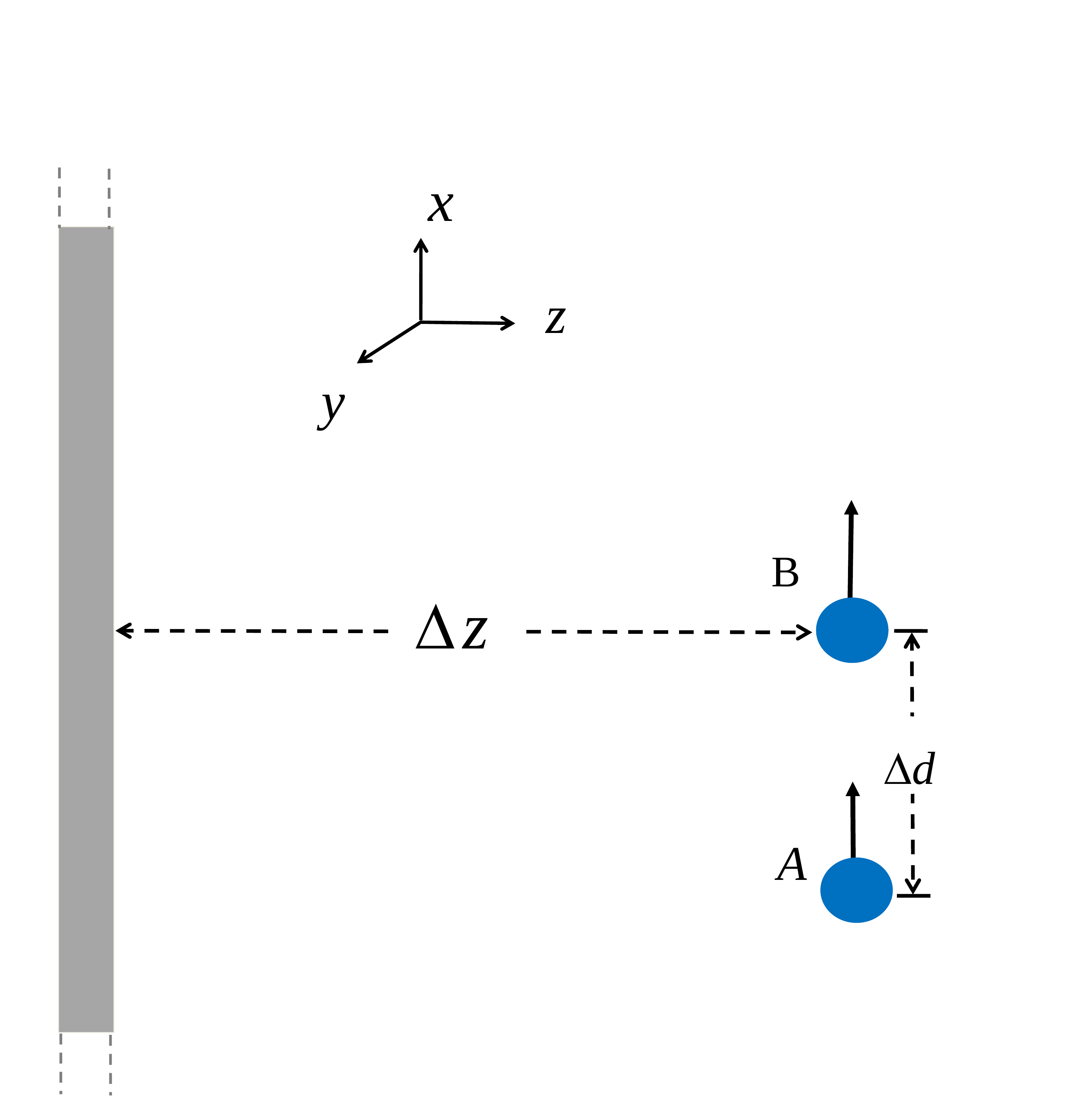}}\qquad\qquad\qquad
 \subfloat[] {\includegraphics[width=0.35\linewidth]{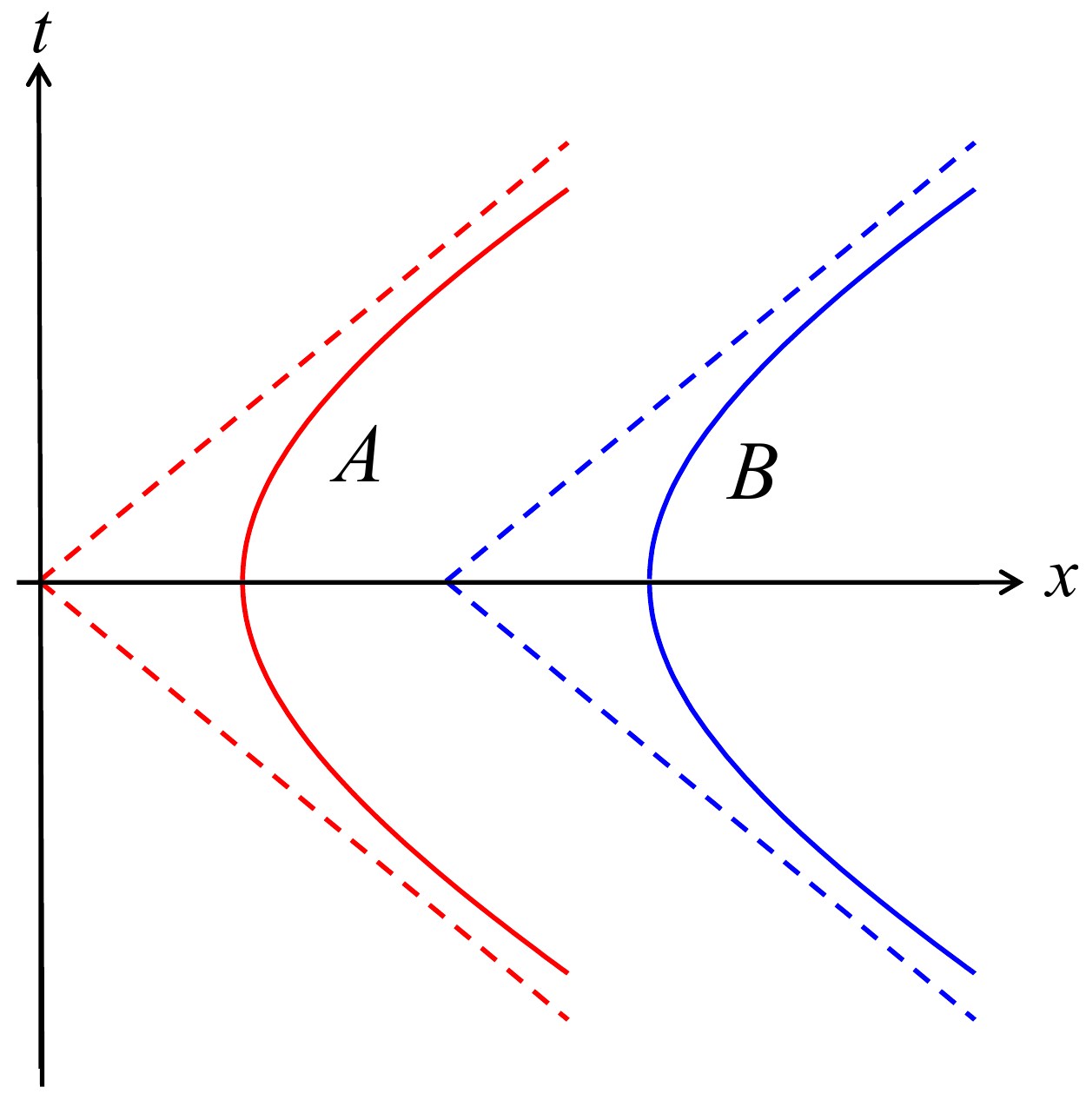}}
  \caption{ In the case of parallel acceleration,  two uniformly accelerated detectors with a separation $\Delta{d}$ are aligned parallel to the
boundary plane at a distance $\Delta {z}$ away from the boundary in (a),  the worldlines of such two accelerated detectors are depicted
respectively in (b). Here, the black arrows indicate the direction  of acceleration.}
  \label{pal}
\end{figure}

Here, we assume that two detectors  are accelerated in parallel with respect to the boundary with
the same magnitude of acceleration (see Fig.~(\ref{pal})), the
corresponding trajectories satisfy~\cite{Salton-Man:2015}
\begin{align}\label{traj-ua}
&x_A:=\{t=a^{-1}\sinh(a\tau_A)\;,~x=a^{-1}\cosh(a\tau_A)\;,~y=0\;,~z=\Delta{z}\}\;,
\nonumber\\
&x_B:=\{t=a^{-1}\sinh(a\tau_B)\;,~x=a^{-1}\cosh(a\tau_B)+\Delta
d\;,~y=0\;,~z=\Delta{z}\}\;,
\end{align}
where $\Delta d$ represents the separation between two detectors, as measured  by an inertial observer at a fixed $x$ (i.e., in the laboratory reference frame).

The transition probabilities can be evaluated  by using
Eq.~(\ref{PD-1}), and the nonlocal correlation term $X$
can be obtained by substituting the Wightman function into
Eq.~(\ref{xxdef}). Here, we use $ X_{((}$ to denote X in the case of the
parallel acceleration where the trajectories of two detectors are
described by Eq. (\ref{traj-ua}).  According to Eq.~(\ref{xxdef}), we have
\begin{align}\label{xxdef-2}
X_{((}=-\lambda^{2}\int_{-\infty}^{\infty} d\tau\int_{-\infty}^{\tau} d \tau' \chi(\tau)\chi(\tau')  e^{-i\Omega( \tau+\tau')}
\Big[W\left( x_A(\tau'),x_B(\tau)\right)+W\left(x_B(\tau'),x_A(\tau)\right)\Big]\;.
\end{align}
Letting $u=\tau$  and $s=\tau-\tau'$, Eq.~(\ref{xxdef-2}) becomes,  after
some algebraic manipulations,
\begin{align}\label{XPP}
X_{((}=- \frac{\lambda^2 a^{2}}{4\pi^2} \int_{-\infty}^{\infty}
du\int_{0}^{\infty} ds f(u,s)\Big[ \frac{1}{f_{\rm{AB}}(u,s)}-
\frac{1}{f_{\rm{AB}}(u,s)+ 4 a^{2} \Delta
z^{2}}+\frac{1}{f_{\rm{BA}}(u,s)}- \frac{1}{f_{\rm{BA}}(u,s)+ 4
a^{2} \Delta z^{2}}\Big]\;,
\end{align}
where
\begin{equation}
f(u,s)=\exp [{(2 {u s-s^{2}-2 u^{2}})}/{2\sigma^2}-i(2u-s)\Omega]\;,
\end{equation}
\begin{equation}
f_{\rm{AB}}(u,s)=2+a^{2} \Delta d^{2}-2 \cosh(a s)+2 a \Delta{d}
\cosh (a u)-2 a \Delta{d}\cosh [a(u-s)]-i \epsilon\;,
\end{equation}
and
\begin{equation}
f_{\rm{BA}}(u,s)=2+a^{2} \Delta d^{2}-2 \cosh(a s)-2 a \Delta{d}
\cosh (a u)+2 a \Delta{d}\cosh [a (u-s)]-i \epsilon\;.
\end{equation}
In principle,  the  transition probabilities $P_D$ of the uniformly accelerated
detectors can be  calculated by using
Eq.~(\ref{PD-1}), and
   the nonlocal correlation term $X_{((}$ can also be  obtained
from
  Eq.~(\ref{XPP}). However,  analytical results are quite difficult to obtain  due to the presence of  the Gaussian switching
  function.  So,  numerical calculations will be
resorted to later on.  The concurrence
Eq.~(\ref{condf}), which quantifies the entanglement harvested by
detectors,  can be evaluated explicitly, once the values of $X_{((}$ and transition
probabilities are known.  Before we show the results of numerical calculations, let us first give the same general analysis for the other two scenarios.

\subsubsection{Anti-parallel acceleration}
\begin{figure}[!htbp]
\centering
\subfloat[]{\includegraphics[width=0.35\linewidth]{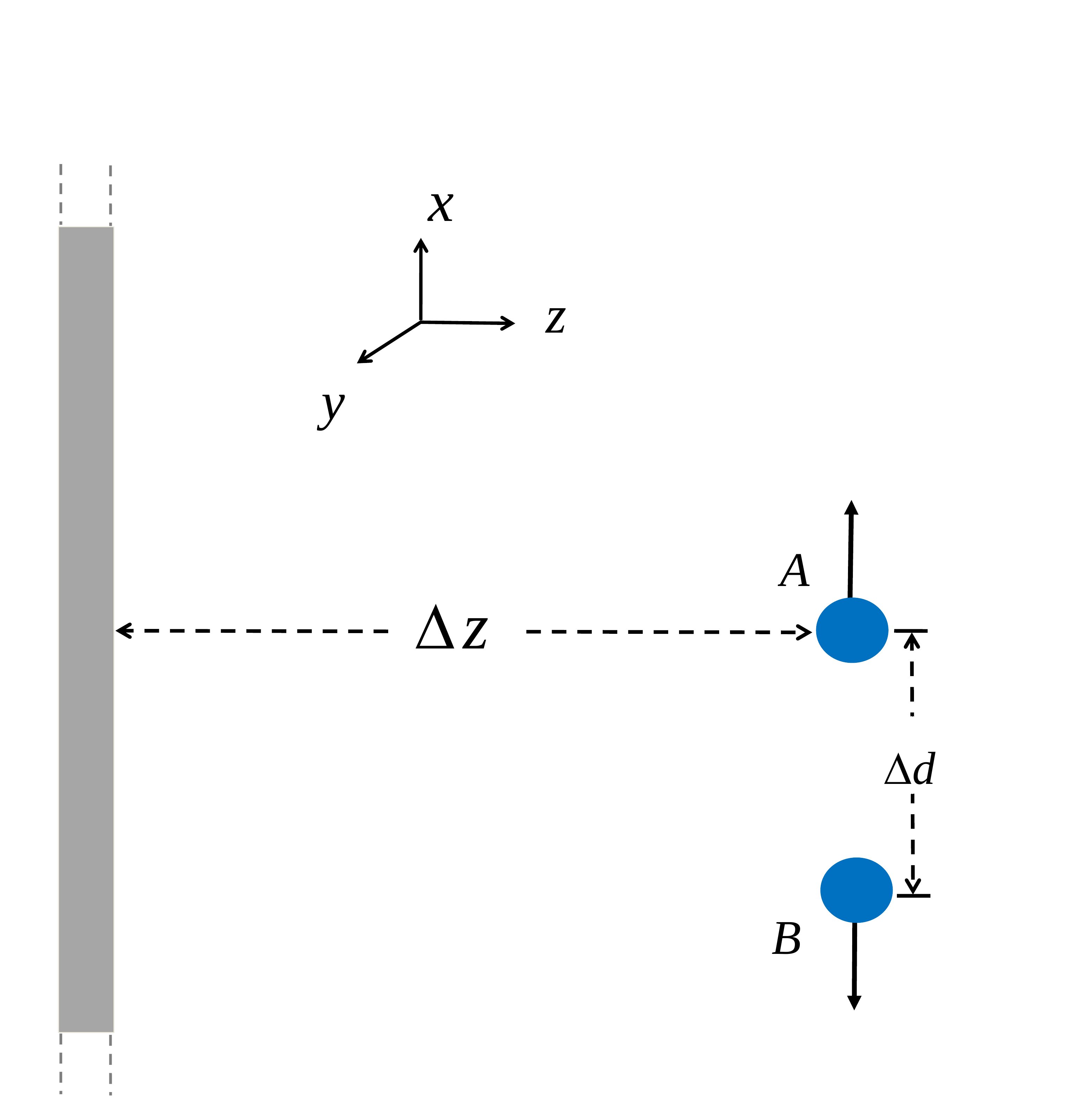}}\qquad\qquad\qquad
 \subfloat[]{\includegraphics[width=0.35\linewidth]{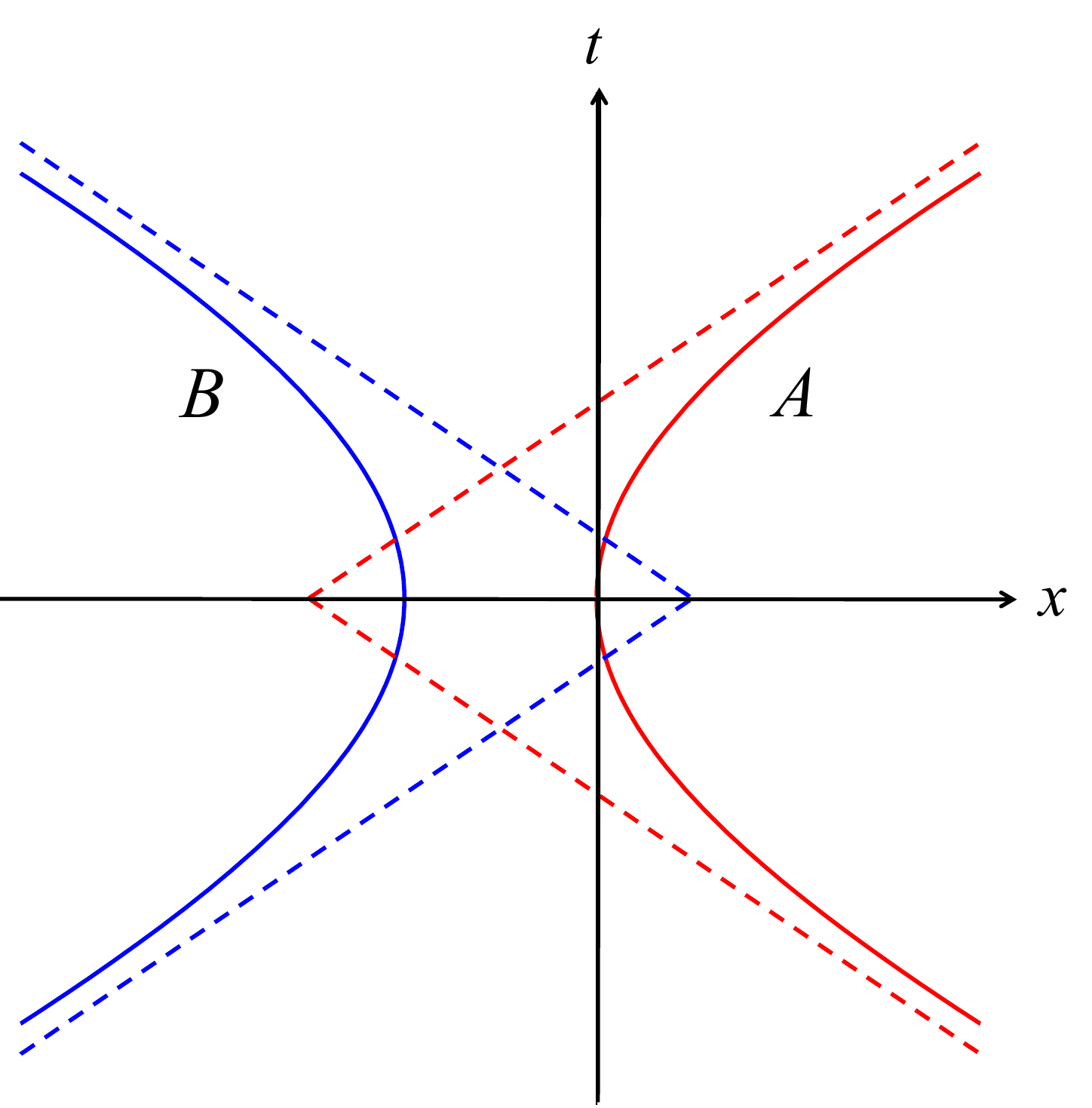}}
\caption{(a) Two  detectors  in anti-parallel acceleration near the
reflecting boundary with a nearest longitudinal shift $\Delta{d}$
along the direction of acceleration, (b) the corresponding
worldlines of two detectors. Black arrows indicate the direction  of acceleration.}\label{anti-par}
\end{figure}

Now, we move on to
the case of two detectors with anti-parallel acceleration along
$x$-axis (see Fig.~(\ref{anti-par})). For convenience, we specify the  spacetime trajectories of
two such detectors in the form as~\cite{Salton-Man:2015}
\begin{align}\label{traj-aa}
&x_A:=\{t=a^{-1}\sinh(a\tau_A)\;,~x=a^{-1}[\cosh(a\tau_A)-1]\;,~y=0\;,~z=\Delta{z}\}\;,
\nonumber\\
&x_B:=\{t=a^{-1}\sinh(a\tau_B)\;,~x=-a^{-1}[\cosh(a\tau_B)-1]-\Delta
d\;,~y=0\;,~z=\Delta{z}\}\;,
\end{align}
where $\Delta d$ denotes the separation between two detectors at the
closest approach (i.e., at the origin of  the time coordinate), as seen by a rest observer at a constant $x$, namely the closest distance
$\Delta{d}$ is independent of the acceleration~\cite{Salton-Man:2015}. It
should be point out that the trajectories Eq.~(\ref{traj-aa}) in
general relax the condition of overlapping apexes shared by four
Rindler wedges of two detectors.

The transition probabilities can be numerically calculated  by using
Eq.~(\ref{PD-1}) as well,  and the nonlocal correlation term $X$, now
represented by $X_{)(}$,
can be obtained by substituting Eq.~(\ref{traj-aa}) into
Eq.~(\ref{xxdef})
\begin{align}
X_{)(}=- \frac{\lambda^2 a^{2}}{2\pi^2} \int_{-\infty}^{\infty} du
\int_{0}^{\infty} ds f(u,s)\left[\frac{1}{g(u,s)}- \frac{1}{g(u,s)+4
a^{2} \Delta{z}^2}\right]\;,
\end{align}
where
\begin{equation}
g(u,s) =2+(2-a \Delta {d})^2 +2 \cosh (a s-2 a u)+(2 a \Delta d-4) [
\cosh (a u)+ \cosh (a u-a s)]-i \epsilon\;.
\end{equation}
Numerically evaluating the transition probability $P_D$
and $X_{)(}$\;,  we can  explore the concurrence
$\mathcal{C}(\rho_{A B})$ to reveal the features of entanglement harvesting, which we will present  in detail later. 

\subsubsection{The acceleration in perpendicular orientations}
\begin{figure}[!htbp]
  \centering
  \subfloat[]{\includegraphics[width=0.35\linewidth]{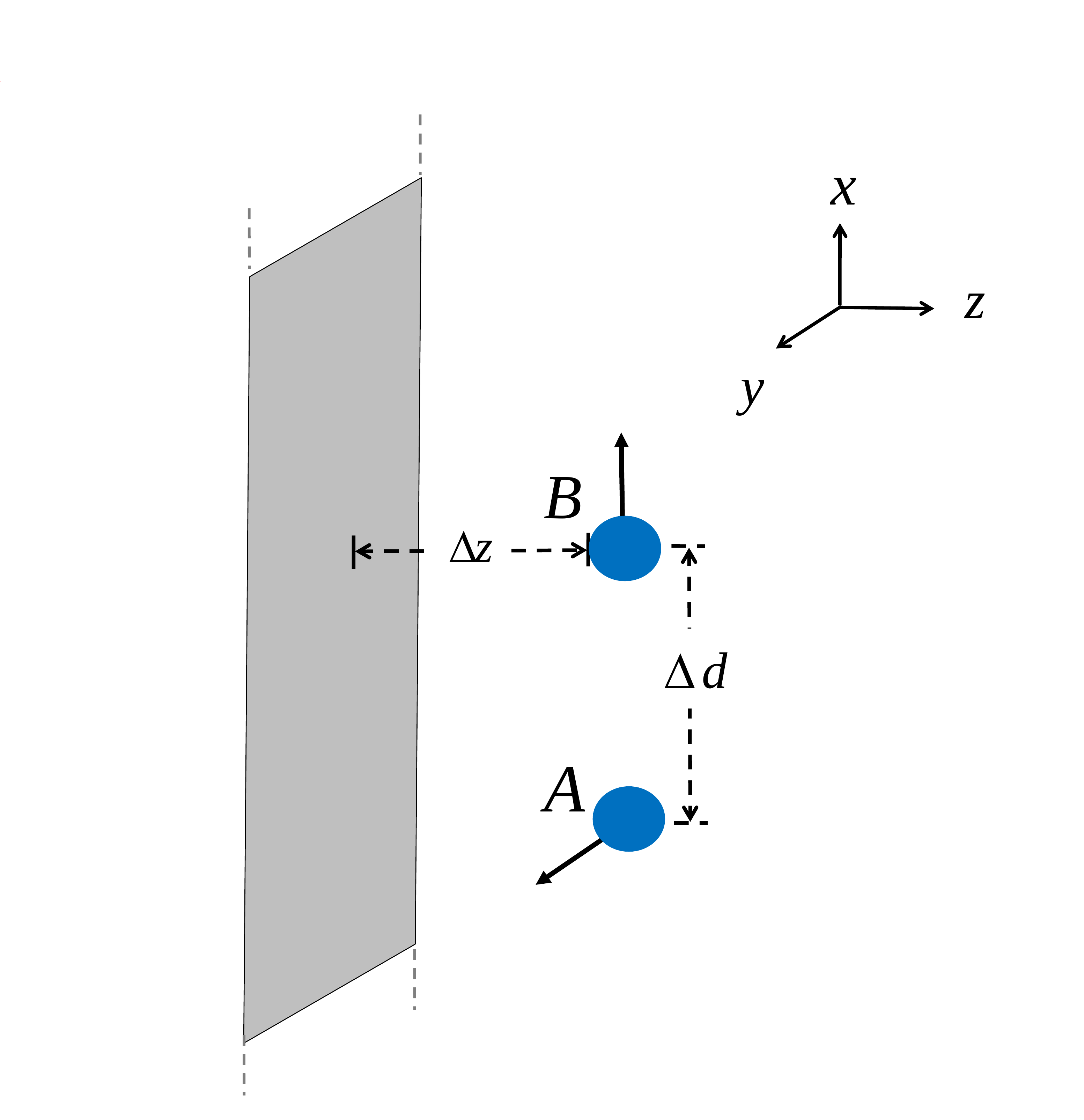}}\qquad\qquad\qquad
   \subfloat[]{\includegraphics[width=0.35\linewidth]{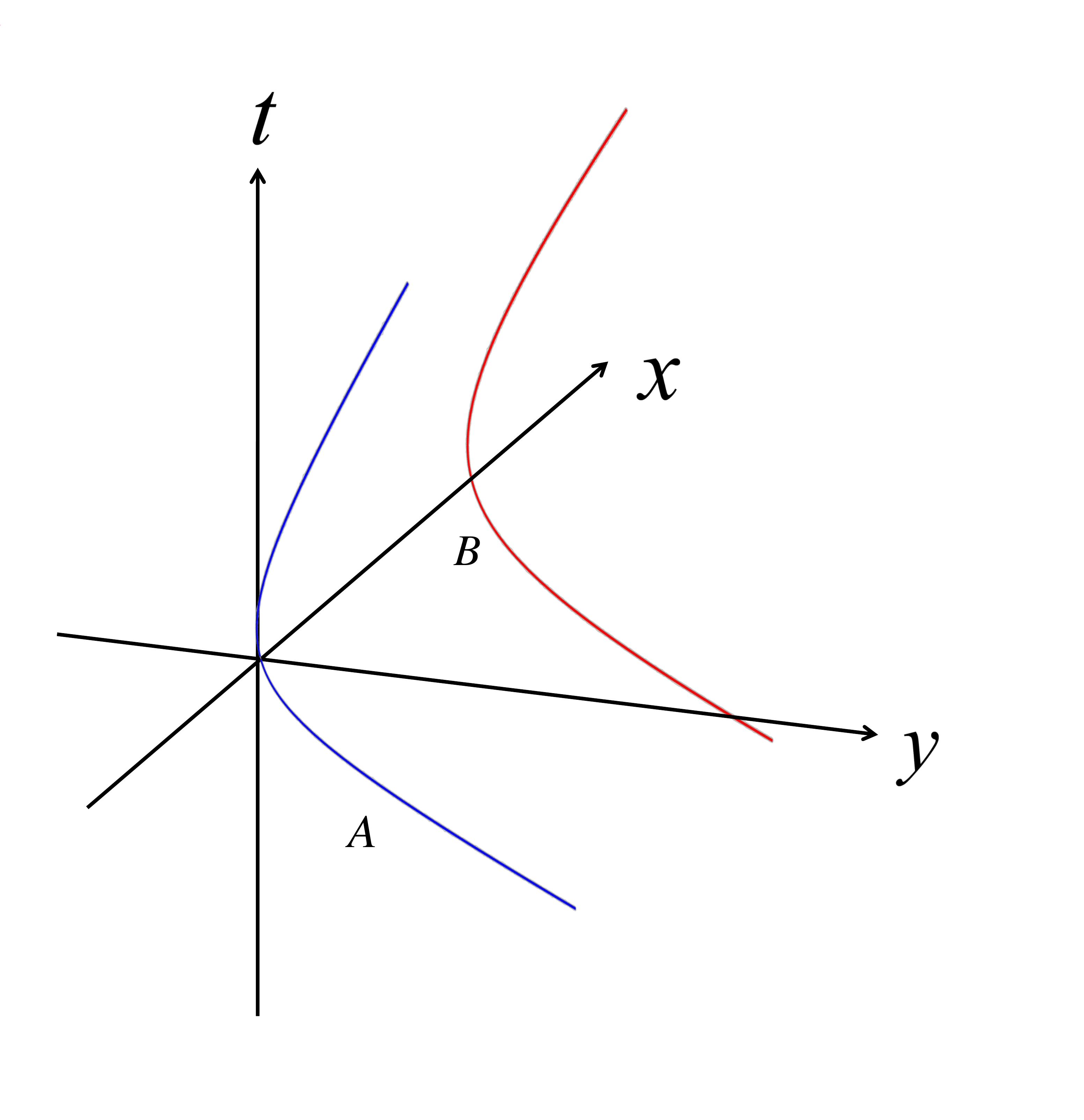}}
  \caption{(a) The plot of two uniformly accelerated detectors with the acceleration in mutual perpendicular directions near the reflecting boundary, (b) the corresponding worldlines of such two detectors.  Black arrows indicate the direction  of acceleration.}
  \label{ver}
\end{figure}

In this subsection, we will consider the case of  two detectors uniformly accelerated  in mutually vertical directions.  Concretely,  we assume that two detectors are uniformly accelerated in a plane parallel to the boundary, where  one detector is accelerated in the  $x$-axis and the other  accelerated in the $y$-axis (see Fig.~(\ref{ver})).
Hence, the spacetime  trajectories of the two perpendicularly accelerated detectors take the following form
\begin{align}\label{traj-va}
&x_A:=\{t=a^{-1}\sinh(a\tau_A)\;,~x=0\;,~y=a^{-1}[\cosh(a\tau_A)-1]\;,~z=\Delta{z}\}\;,
\nonumber\\
&x_B:=\{t=a^{-1}\sinh(a\tau_B)\;,~x=a^{-1}[\cosh(a\tau_B)-1]+\Delta{d}\;,~y=0\;,~z=\Delta{z}\}\;.
\end{align}
Similarly, a substitution of  Eq.~(\ref{traj-va}) into Eq.(\ref{xxdef}) yields  the
nonlocal correlation term $X$, denoted by $X_{\perp}$ in the present case
\begin{equation}
X_{\perp}=- \frac{\lambda^2 a^{2}}{4\pi^2} \int_{-\infty}^{\infty}
du \int_{0}^{\infty} ds f(u,s)\left[\frac{1}{h_{\rm{AB}}(u,s)}-
\frac{1}{h_{\rm{AB}}(u,s)+ 4 a^{2} \Delta
z^{2}}+\frac{1}{h_{\rm{BA}}(u,s)}- \frac{1}{h_{\rm{BA}}(u,s)+ 4
a^{2} \Delta z^{2}}\right]
\end{equation}
where
\begin{align}
h_{\rm{AB}}(u,s) =3+(a \Delta{d}-1)^2-2 \cosh[a(u-s)]-2(1-a \Delta
{d}) \cosh [a u]+2 \sinh (a u) \sinh [a (u-s)]-i \epsilon\;,
\end{align}
and
\begin{align}
h_{\rm{BA}}(u,s) =3+(a \Delta{d}-1)^2-2 \cosh(a u)-2(1-a \Delta d)
\cosh [a (u-s)]+2 \sinh (a u) \sinh [a (u-s)]-i \epsilon\;.
\end{align}
Now with the formulae needed for examining the entanglement harvesting by two detectors uniformly accelerated in  all the three different scenarios,
we are  to show the results of our numerical calculations below.

\subsubsection{The numerical result  and cross comparison  }
\begin{figure}[!htbp]
\centering
\subfloat[$a\sigma=0.10$]{\includegraphics[width=0.32\linewidth]{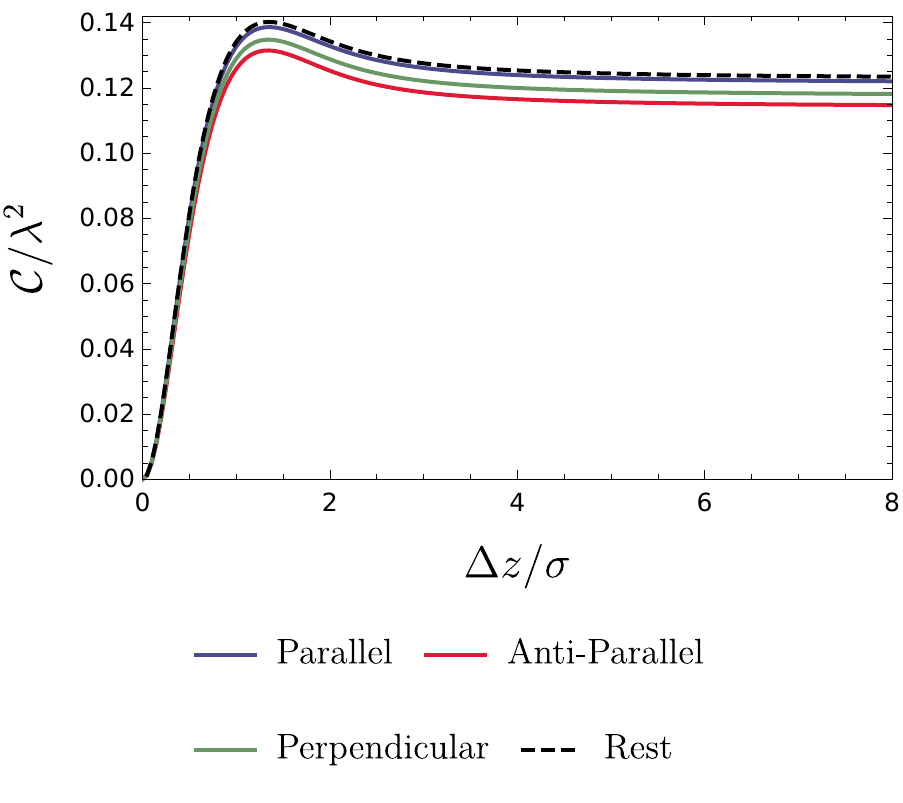}}\;\;
 \subfloat[$a\sigma=0.50$]{\includegraphics[width=0.32\linewidth]{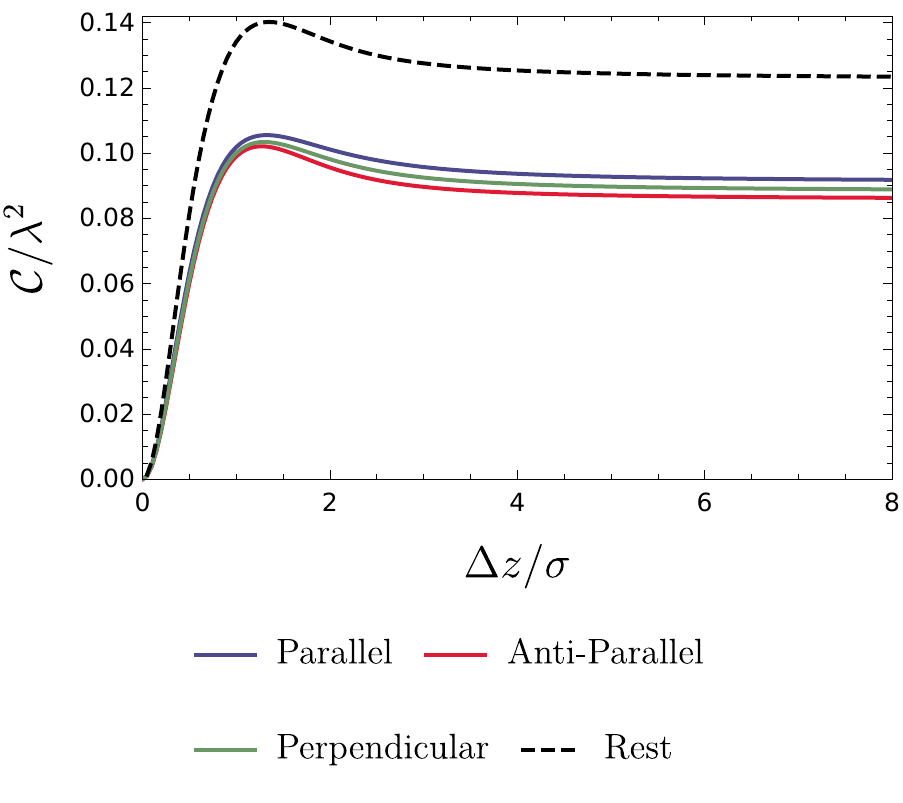}}\;\;\subfloat[$a\sigma=1.00$]{\includegraphics[width=0.32\linewidth]{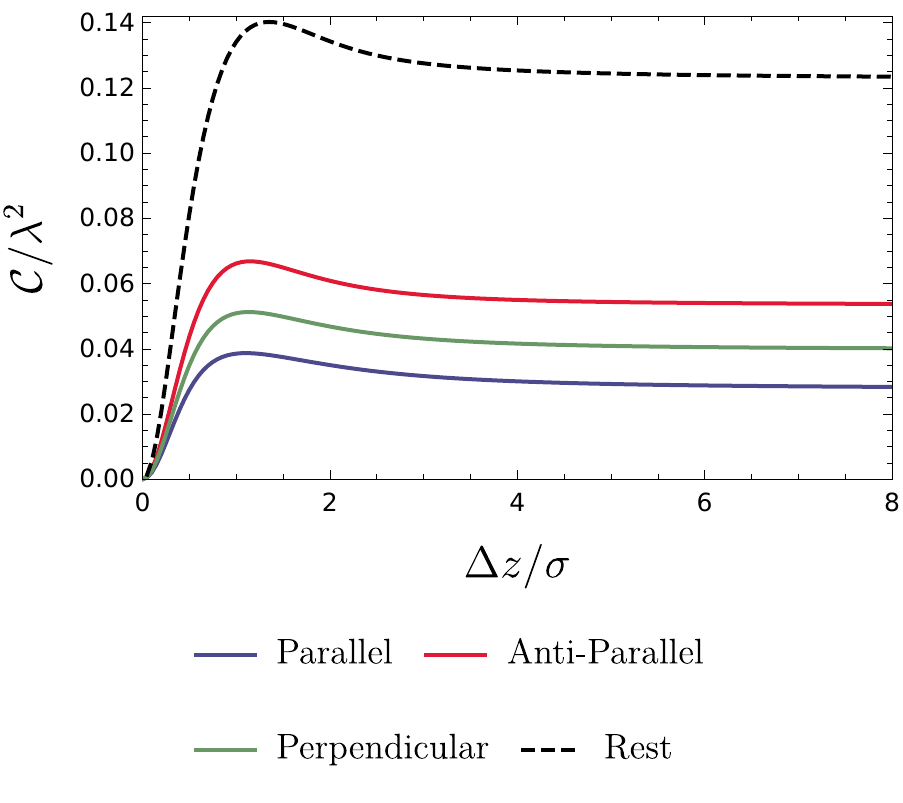}}
\caption{The concurrence is plotted as a function of
  $\Delta{z}/\sigma$  with  fixed $\Omega\sigma=0.10$,
  $\Delta{d}/\sigma=1.00$,  and $a\sigma=\{0.10,0.50,1.00\}$ from left to right. In the limit of $\Delta{z}\rightarrow0$, the concurrence $\mathcal{C}(\rho_{A B})$ in all acceleration scenarios will approach zero regardless of the values of  acceleration $a$.  Obviously, there is a peak of concurrence near $\Delta{z}/\sigma \sim1$. }\label{condzvsa}
 \end{figure}

\begin{figure}[!htbp]
\centering
\subfloat[$\Delta{d}/\sigma=0.50$]{\includegraphics[width=0.32\linewidth]{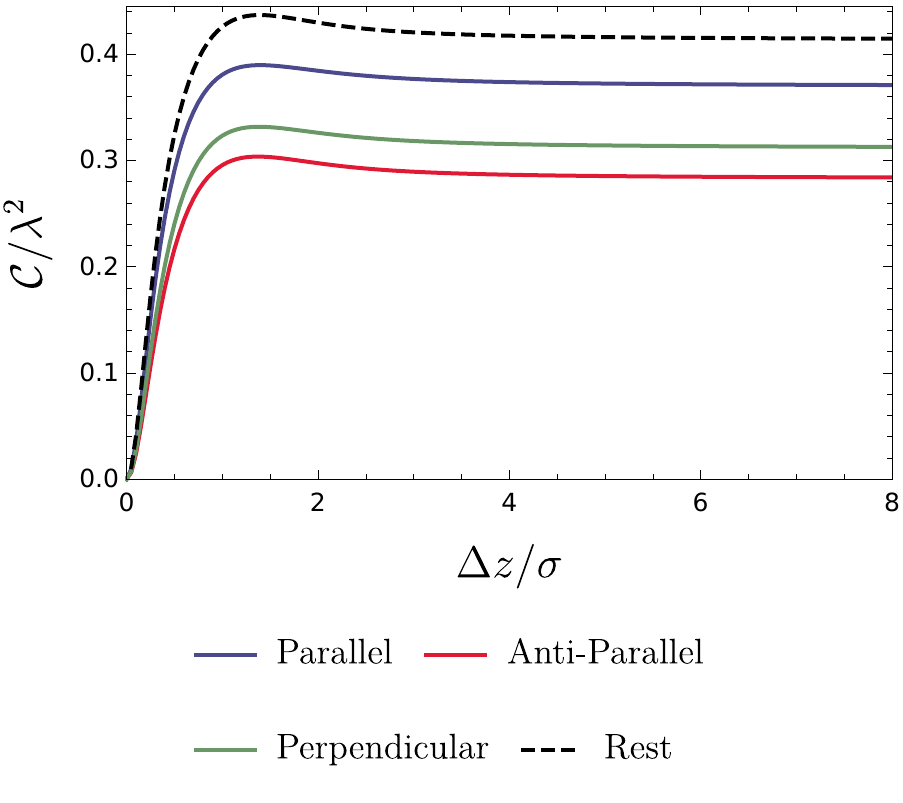}}\;\;
 \subfloat[$\Delta{d}/\sigma=0.80$]{\includegraphics[width=0.32\linewidth]{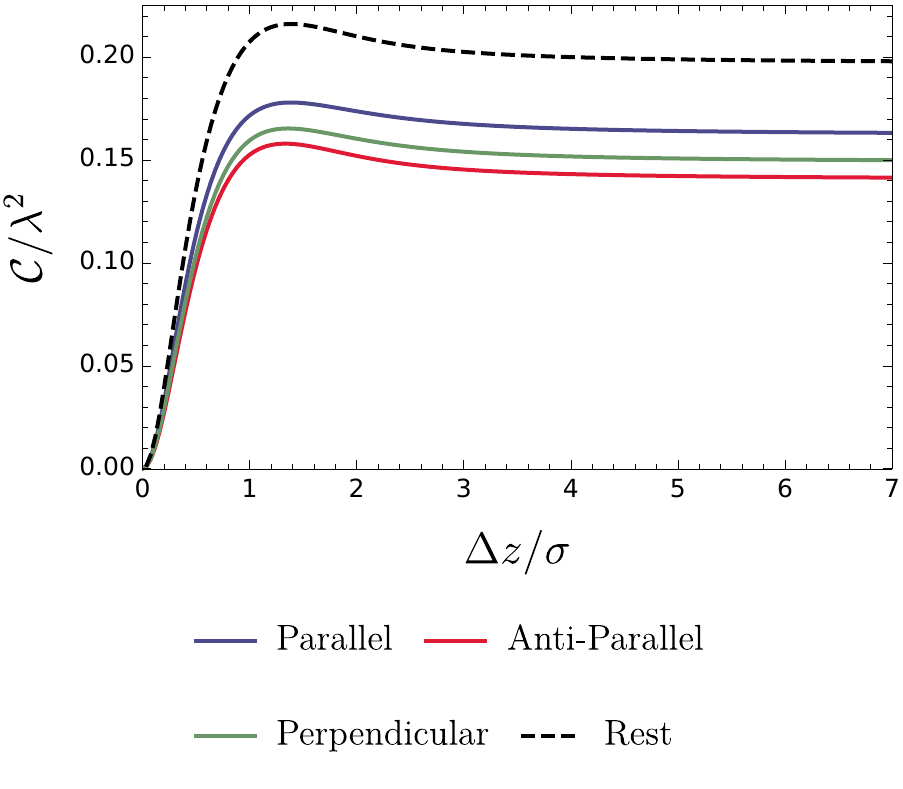}}\;\;\subfloat[$\Delta{d}/\sigma=1.50$]{\includegraphics[width=0.32\linewidth]{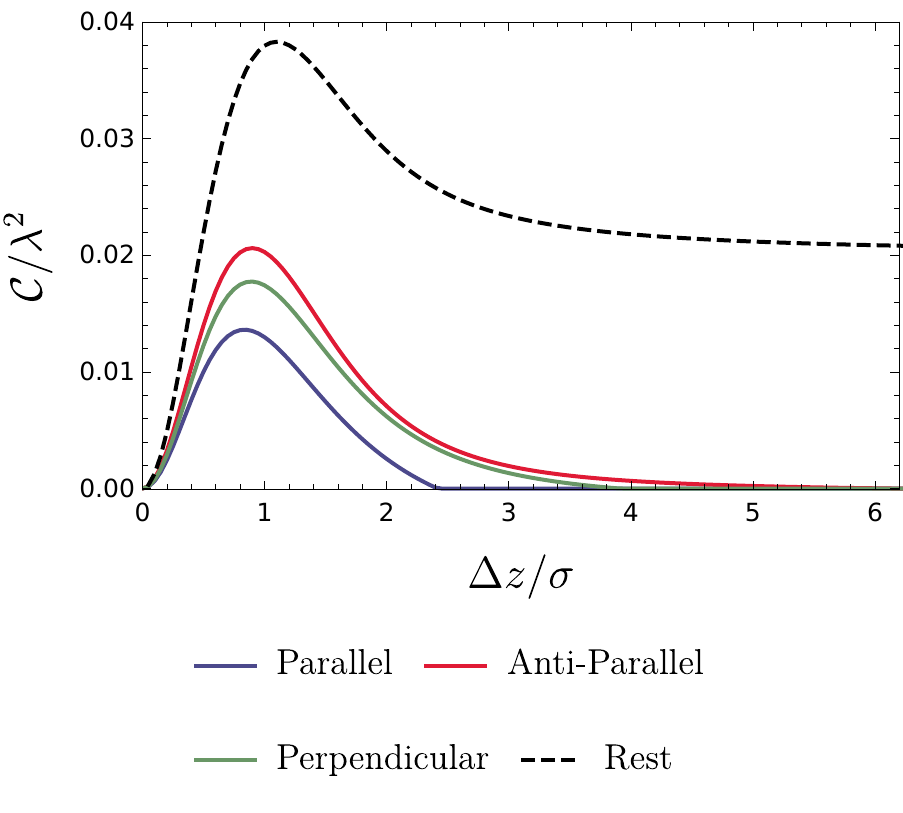}}
\caption{The concurrence is plotted as a function of
  $\Delta{z}/\sigma$ for various separation  $\Delta{d}/\sigma=\{0.50,0.80,1.50\}$. Here, we have set $\Omega\sigma=0.10$ and  $a\sigma=0.50$. In the limit of $\Delta{z}\rightarrow0$, the concurrence $\mathcal{C}(\rho_{A B})$ approaches zero regardless of the value of the separation $\Delta{d}$. }\label{condzvsd}
 \end{figure}

 We  now turn to explore the entanglement harvesting phenomenon for two detectors in parallel, anti-parallel and mutually perpendicular
acceleration via numerical evaluation.We begin by plotting, in Figs.~(\ref{condzvsa}) and (\ref{condzvsd}), the concurrence  as a function of
the distance between the boundary and the detectors, $\Delta z$, with other parameters of the system fixed at certain values for all three acceleration scenarios. As we can
see, the distance from the boundary $\Delta{z}/\sigma$ significantly
influences the harvested entanglement.  Particularly, similar to that in the situation of inertial detectors at rest, the entanglement harvesting  is greatly inhibited when the detectors are close to the boundary ($\Delta{z}/\sigma\ll1$). It is worth pointing out that the concurrence  asymptotically  approaches that without a boundary as  the distance between the detectors and boundary grows very large (i.e., $\Delta{z}/\sigma\gg1$),  as a result of the fact that the Wightman function reduces to its free-space form in the limit of $\Delta{z}\rightarrow\infty$.  In fact, this is valid irrespective of whether the detector is accelerating or not. Again similar to that in the situation of inertial detectors at rest, there also exists, for the same reason, a peak of the extracted entanglement approximately at the position where $\Delta{z}$ is comparable to the duration parameter $\sigma$, for both fixed $\Delta{d}/\sigma$ and $a\sigma$ in all three scenarios. This phenomenon of  entanglement enhancement by a boundary  was similarly  found in Ref.~\cite{Cheng:2018} where entanglement dynamics of uniformly accelerated polarizable atoms coupled with electromagnetic fields in the presence of a reflecting boundary was examined. Here, the general inhibition of entanglement harvesting  and the peak occurrence  of the harvested entanglement that arise from the effects of the boundary still remain regardless of whether  two detectors are accelerated or not. Interestingly, the location of the peak, as shown in Figs.~(\ref{condzvsa}) and (\ref{condzvsd}), is almost independent of the acceleration scenarios, although the  magnitude of the peak is noticeably affected by the acceleration scenario.  In general, the parallel acceleration case seems to have a comparatively large peak for a small acceleration or  separation between two detectors, while  for a large acceleration or separation between two detectors  the anti-parallel acceleration case takes the place.


Now let us turn our attention to a cross comparison of the entanglement harvesting in various scenarios.
From Fig.~(\ref{condzvsa}) and Fig.~(\ref{condzvsd}), one can find that the issue of which acceleration scenario is better in terms of entanglement harvesting crucially
 depends on the value of the acceleration $a\sigma$ or the detectors' separation $\Delta{d}/\sigma$. Since the detectors' thermalization that arises from the effect of  acceleration would generally inhibit the extraction of entanglement, thus the  inertial detectors at rest are likely to harvest more entanglement than those in uniform acceleration scenarios at the same distance $\Delta{z}$ from the boundary with not too large energy gap $\Omega$.   Among  three acceleration scenarios,  the parallel acceleration case is  likely to harvest more entanglement for a small acceleration ($a\sigma<1$ and $a\Delta{d}\ll1$). However, in contrast,  the  anti-parallel acceleration extracts more entanglement for a large acceleration ($a\sigma\gg1$). Physically, these features can be understood as follows.   For  a small acceleration, the  average detectors' separation during the interaction with fields characterized by the duration parameter $\sigma_D$  is the dominant factor in the  non-local correlation term $X$.  The smaller the separation, the large  the  $|X|$. As a result, comparatively  more entanglement is extracted for the parallel acceleration due to the fact that two  detectors in parallel acceleration  has a smaller average detectors' separation (i.e., they  spend  more time closer to one another). This is consistent with what happens for the inertial detectors.
  But, for a large acceleration, the decrease of $|X|$ becomes much more complicated  as now it is a consequence of the intertwined effects
from $a$ and $\Delta{d}$,  so that
  the effect of acceleration dominates in the non-local correlation term $X$ and overweighs that of the detectors' separation,
 resulting in the anti-parallel acceleration scenario harvesting more entanglement although the parallel acceleration case still has a smaller average detectors' separation.

\begin{figure}[!htbp]
\centering
\subfloat[$\Delta{d}/\sigma=0.20$]{\label{comp-a11}\includegraphics[width=0.32\linewidth]{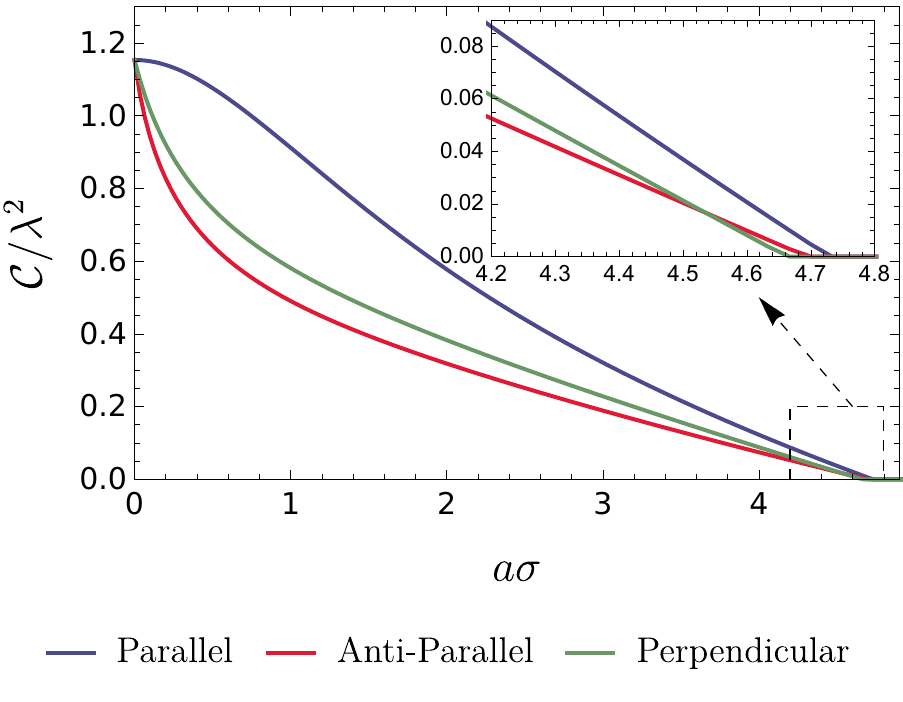}}\quad
 \subfloat[$\Delta{d}/\sigma=0.50$]{\label{comp-a22}\includegraphics[width=0.32\linewidth]{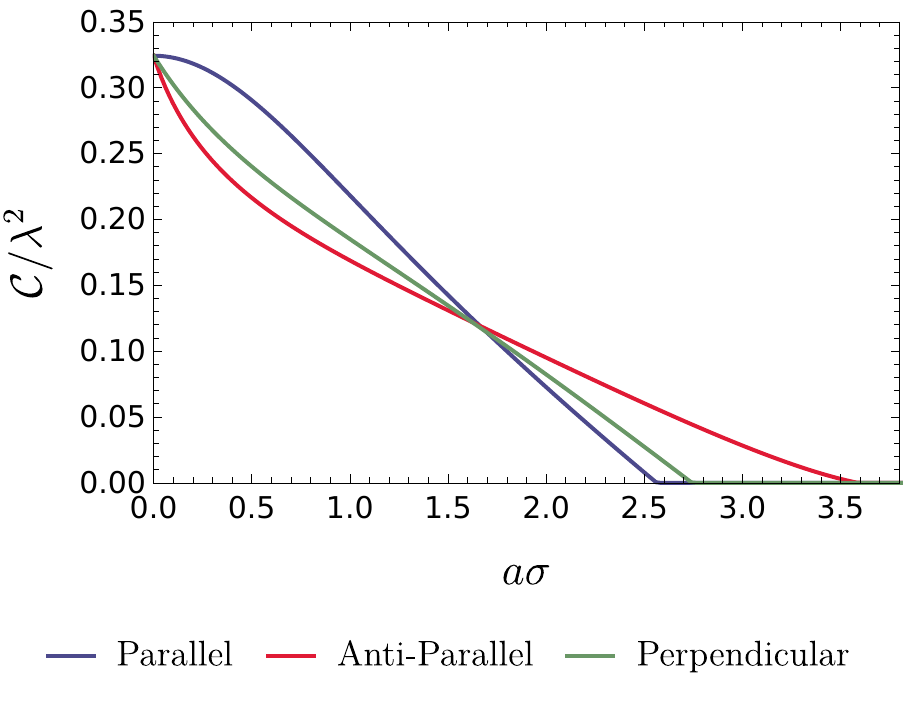}}\quad
 \subfloat[$\Delta{d}/\sigma=1.00$]{\label{comp-a33}\includegraphics[width=0.32\linewidth]{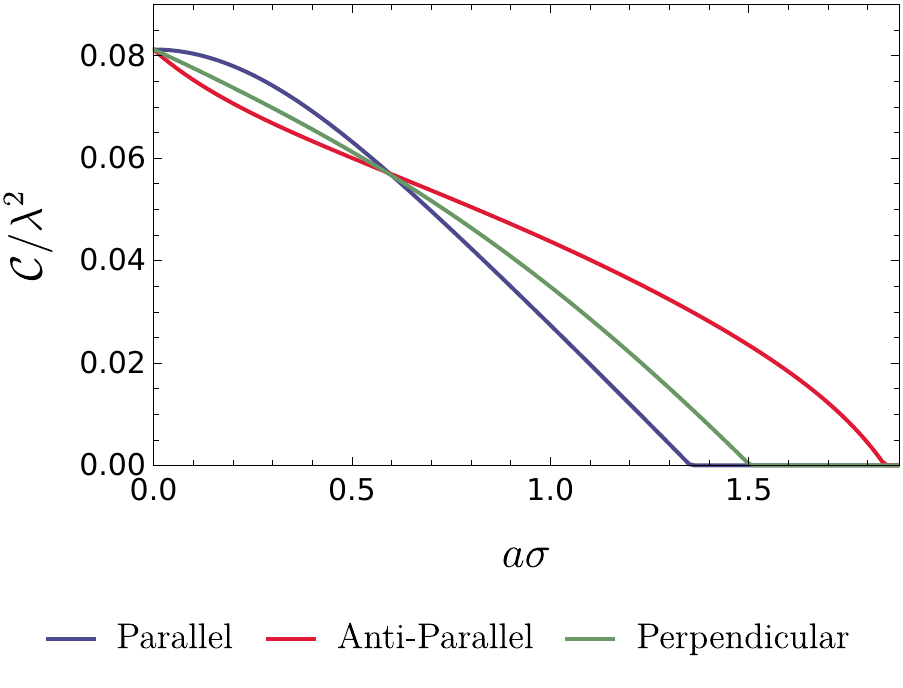}}
  \caption{The concurrence $\mathcal{C}/\lambda^2$  vs  $a\sigma$ for  $\Delta{d}/\sigma=\{0.20,0.50,1.00\}$ in the order from the left-to-right with fixed $\Delta{z}/\sigma=0.50$ and $\Omega\sigma=0.10$. In each plot, the different colored solid lines correspond to the  parallel,  anti-parallel and mutually perpendicular acceleration respectively. When $\Delta{d}/\sigma$ is not too small,  there seems to exist an intersection of the three curves at a nonzero $a\sigma$.  However, for small $\Delta{d}/\sigma$, such intersecting behavior is no longer apparent (e.g., the plot of $\Delta{d}/\sigma=0.20$). }\label{comp-a}
  \end{figure}

\begin{figure}[!htbp]
\centering
\subfloat[$a\sigma=0.10$]{\label{comp-d11}\includegraphics[width=0.32\linewidth]{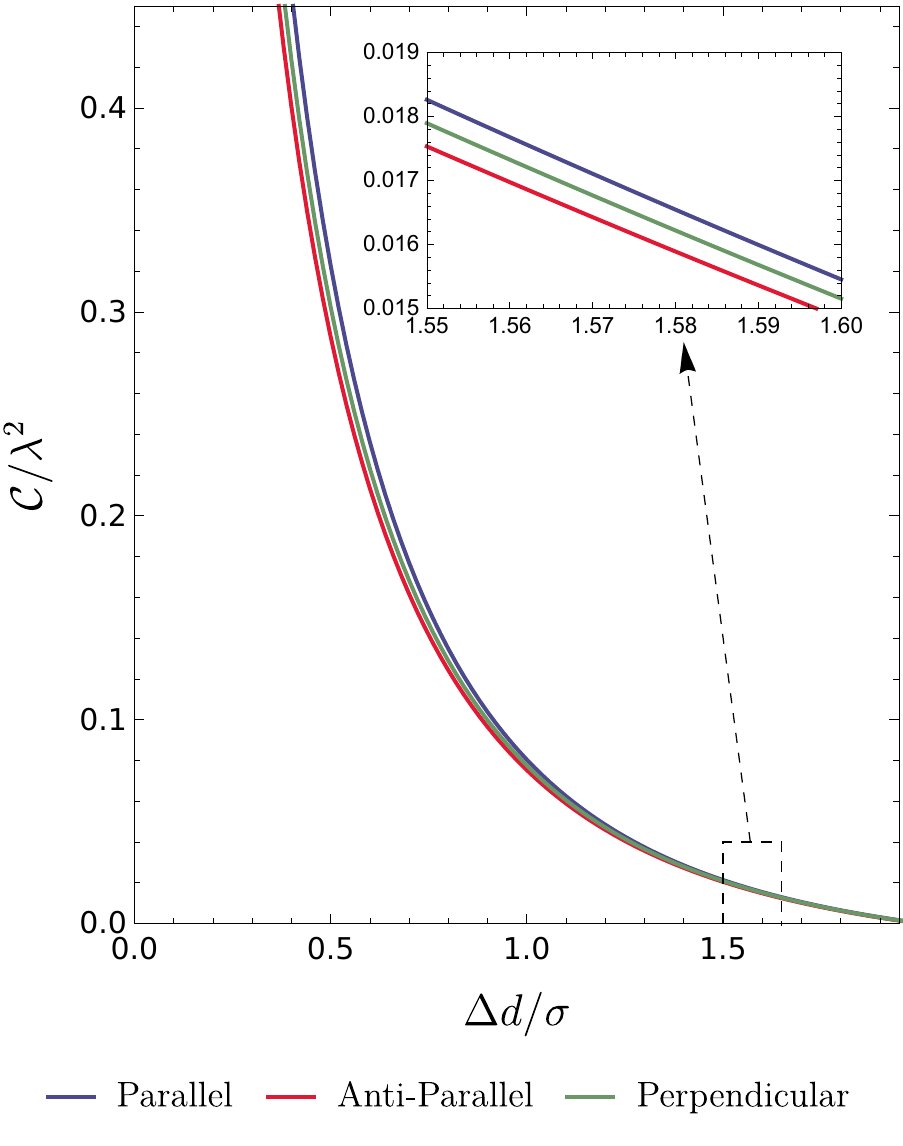}}\quad
 \subfloat[$a\sigma=0.50$]{\label{comp-d22}\includegraphics[width=0.32\linewidth]{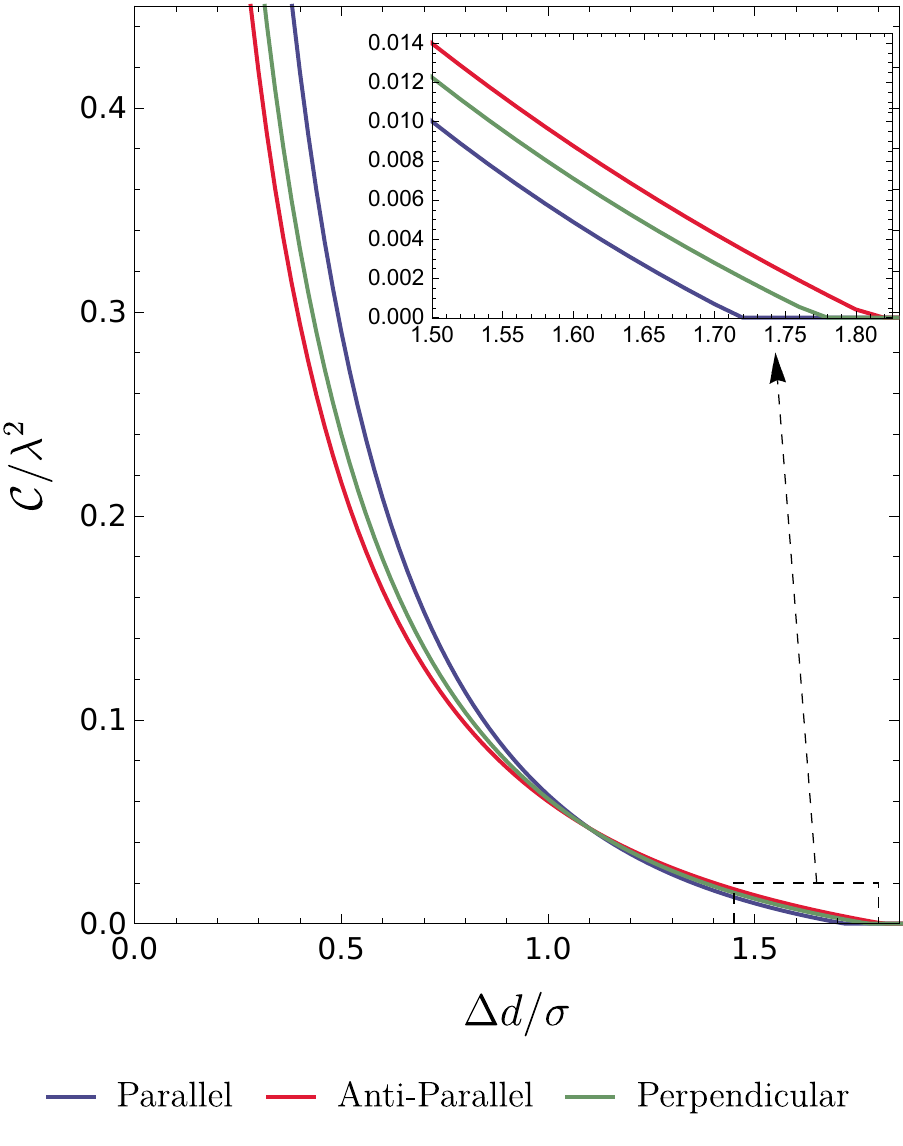}}\quad
 \subfloat[$a\sigma=1.00$]{\label{comp-d33}\includegraphics[width=0.32\linewidth]{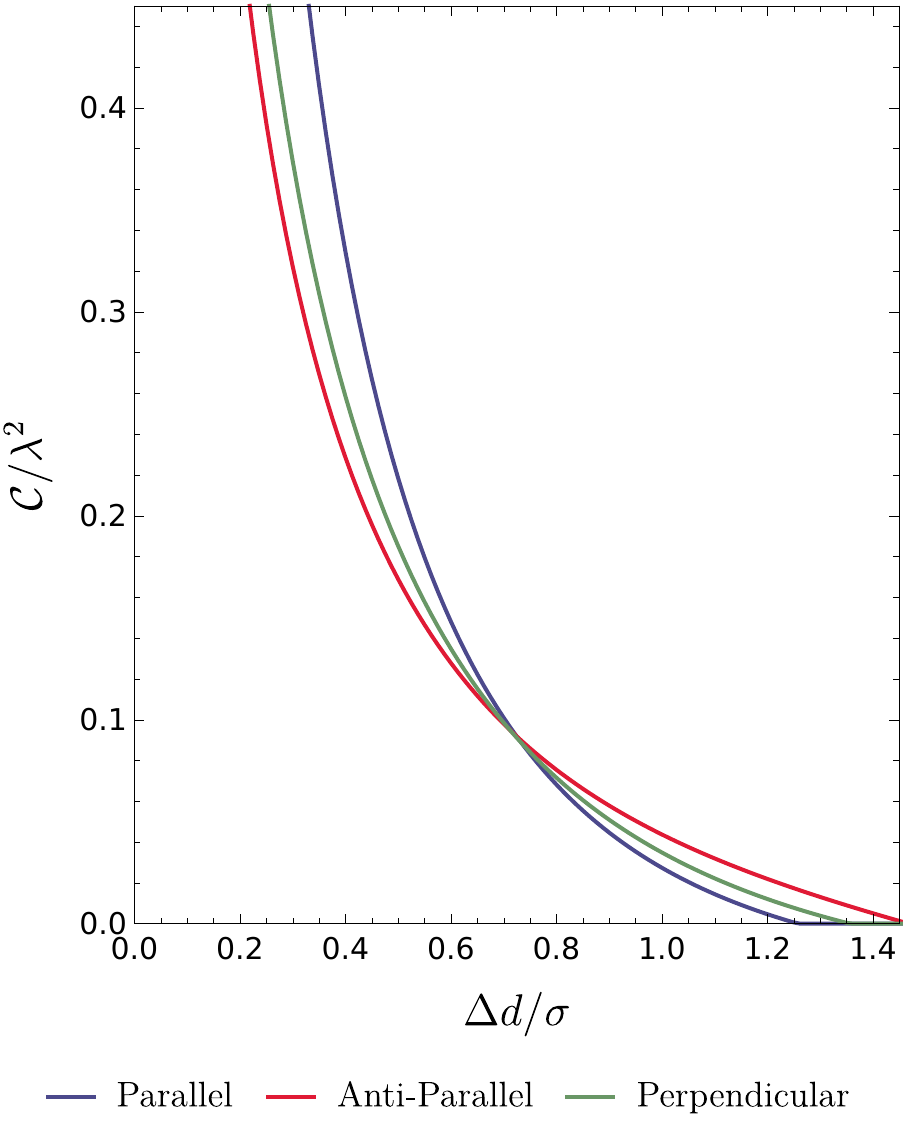}}
  \caption{The plots of the concurrence vs $\Delta{d}/\sigma$ for $a\sigma=\{0.10,0.50,1.00\}$ in the left-to-right order  with fixed $\Delta{z}/\sigma=0.50$ and $\Omega \sigma=0.10$. In each plot, a comparison of different scenarios is implemented via different colored solid curves. }\label{comp-d}
\end{figure}

  For a more thorough analysis on the difference in three acceleration scenarios, we plot  the concurrence as a function of acceleration with different fixed $\Delta{d}/\sigma$ to compare the influence of acceleration for all three acceleration scenarios in Fig~(\ref{comp-a}), and that of the detectors' separation with different fixed acceleration to compare the influence of the detectors' separation in Fig.~(\ref{comp-d}).

As shown in Fig.(\ref{comp-a})/(\ref{comp-d}), the harvested
entanglement in general decreases as either the  acceleration $a$
or  the detectors' separation $\Delta{d}$ increases regardless of
the acceleration scenario. Similar conclusions have also  been reached  in the  case of the circular motion with nonlinear acceleration~\cite{Zhjl:2020}. Interestingly,   as the acceleration increases, the
 concurrence  for the
parallel acceleration case decreases faster than that for the anti-parallel
acceleration and  perpendicular acceleration cases when $\Delta{d}/\sigma$ is not too small (e.g.,
$\Delta{d}/\sigma=0.50$ or $1.00$ in Fig.(\ref{comp-a})).  While for a small
$\Delta{d}/\sigma$, the concurrence for the parallel acceleration is
generally  larger than that for other acceleration scenarios
(see $\Delta{d}/\sigma=0.20$ in Fig.(\ref{comp-a})). Therefore, when
$\Delta{d}/\sigma$ is not too small, the detectors in parallel-acceleration
 comparatively extract more entanglement from the
fields over a certain range of relatively small values of $a\sigma$.  But the harvesting-achievable range
of $a\sigma$, in which the entanglement can be harvested successfully,
may be shorter in the parallel-acceleration case  than that  in the
anti-parallel acceleration and  perpendicular acceleration cases.
Another interesting feature is that for a very large acceleration, entanglement harvesting no longer  occurs although the concrete value of acceleration when the harvesting ceases depends on the acceleration scenario.
Qualitatively similar conclusions can also be drawn from Fig.~(\ref{comp-d}) of the role of detectors' separation
$\Delta{d}/\sigma$ in entanglement harvesting with  $a\sigma$ fixed at certain values.

\begin{figure*}[!htbp]
\centering
\subfloat[$a\sigma=0.10$]{\label{ddethvsz11}\includegraphics[width=0.32\linewidth]{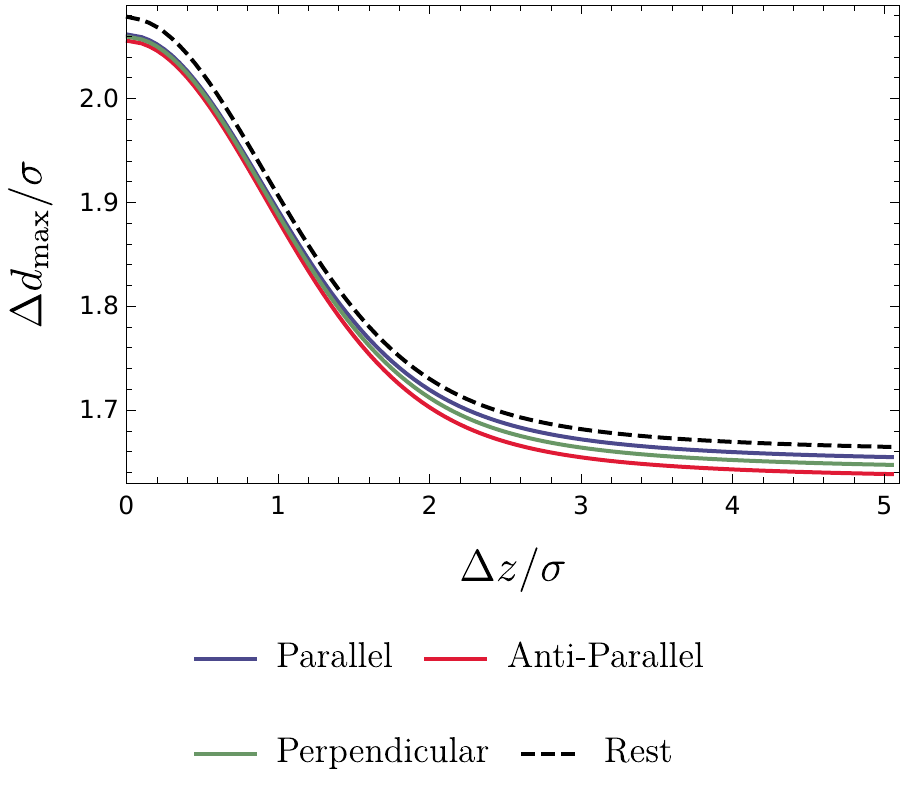}}\quad
 \subfloat[$a\sigma=0.50$]{\label{ddethvsz22}\includegraphics[width=0.32\linewidth]{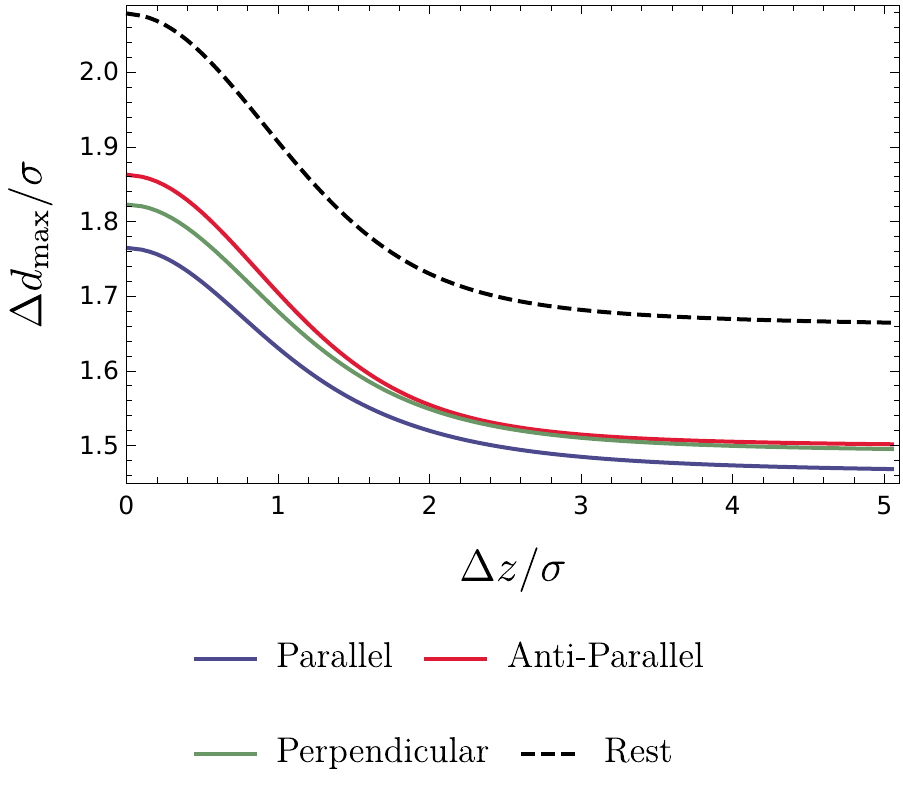}}\quad
 \subfloat[$a\sigma=1.00$]{\label{ddethvsz33}\includegraphics[width=0.32\linewidth]{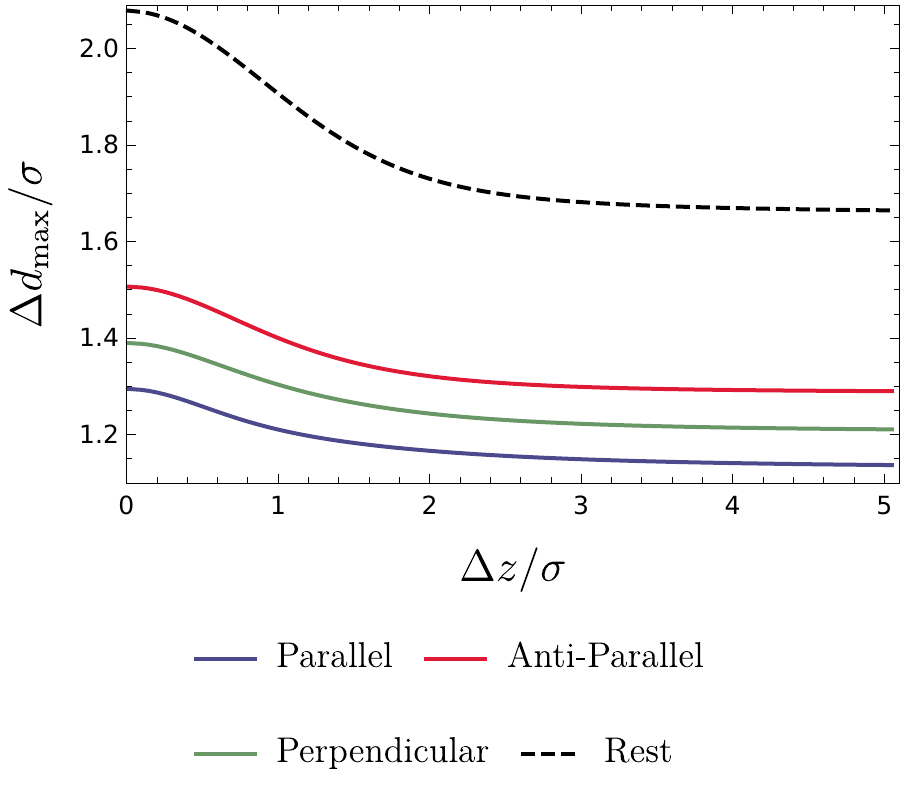}}
\caption{The  separation, $\Delta{d}_{\rm{max}}$, between two detectors when entanglement harvesting  does not occur is plotted as
a function of the distance between detectors and boundary for the
three acceleration scenarios. Here, we have set $\Omega\sigma=0.10$
and $a\sigma=\{0.10\;,0.50\;,1.00\}$ in the left-to-right
order, and the dashed curve denotes the case two rest detectors.}\label{ddethvsz}
\end{figure*}

\begin{figure*}[!htbp]
\centering
\subfloat[$\Delta{d}/\sigma=0.20$]{\label{adethvsz11}\includegraphics[width=0.32\linewidth]{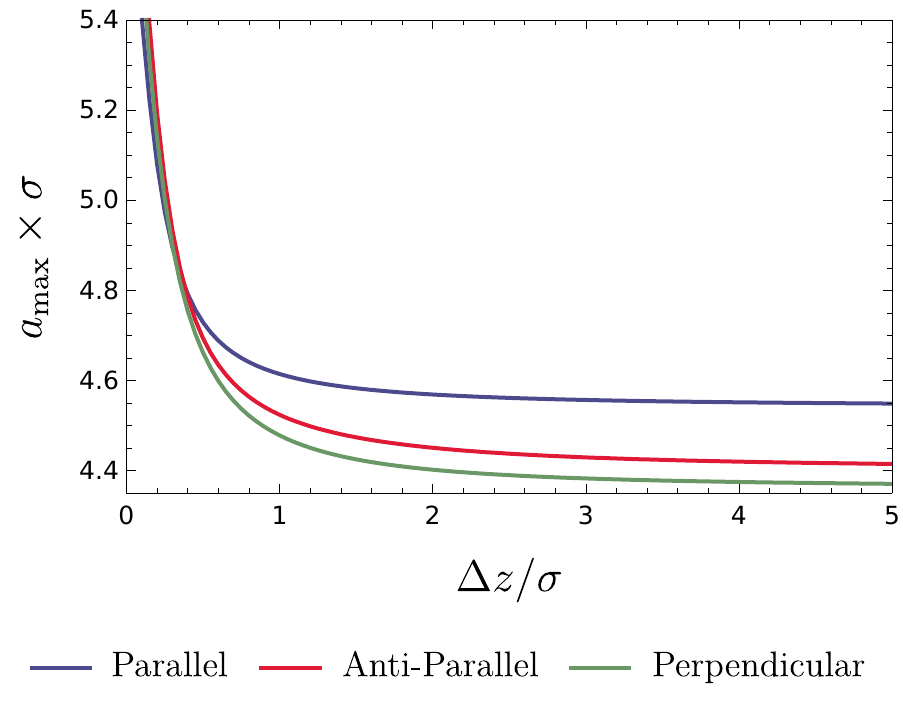}}\quad
 \subfloat[$\Delta{d}/\sigma=0.50$]{\label{adethvsz22}\includegraphics[width=0.32\linewidth]{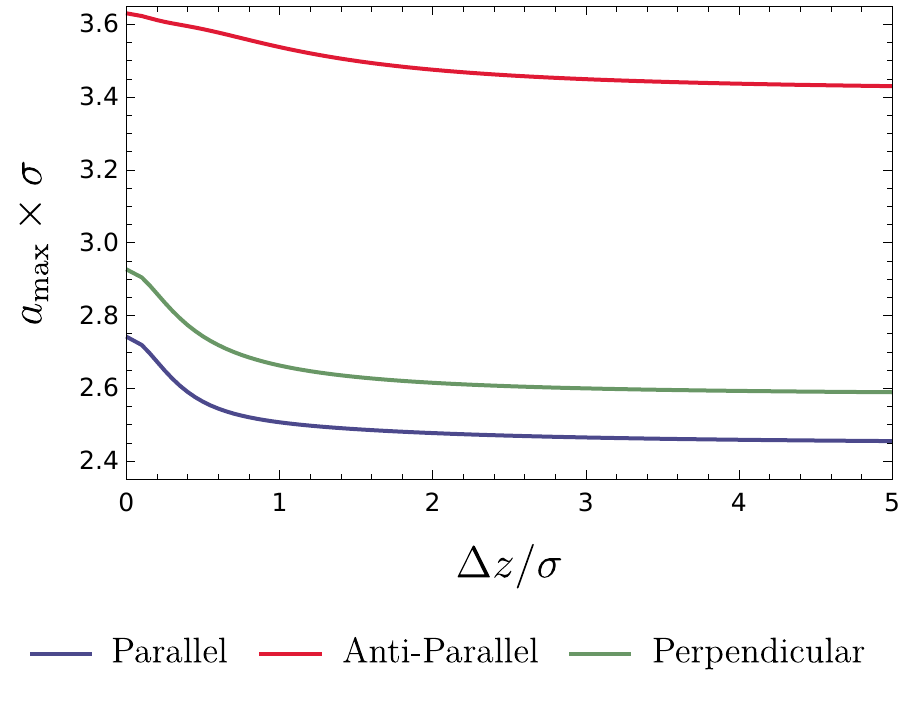}}\quad
 \subfloat[$\Delta{d}/\sigma=1.00$]{\label{adethvsz33}\includegraphics[width=0.32\linewidth]{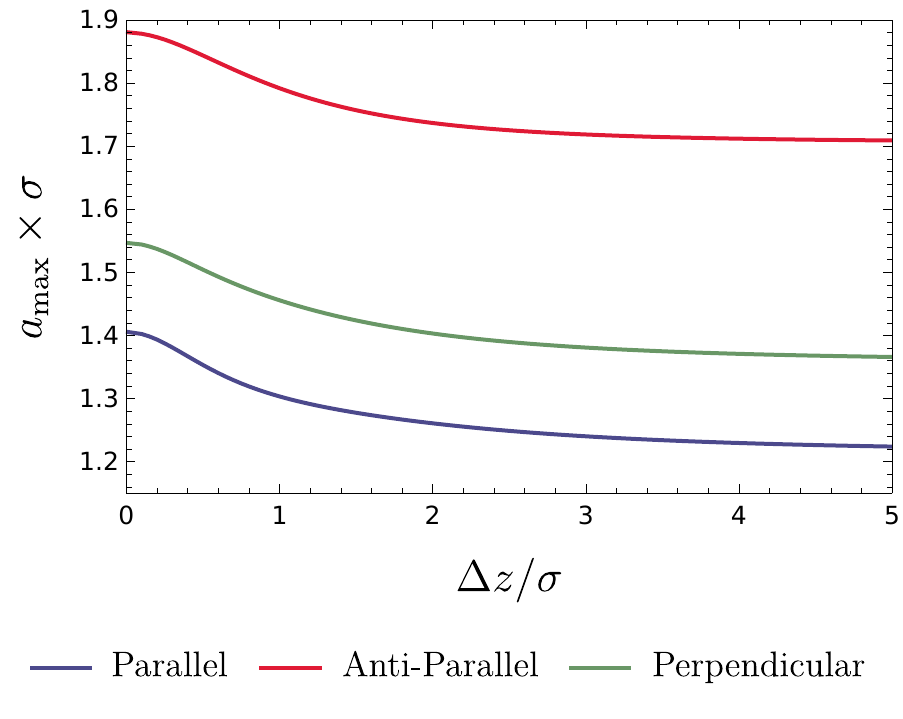}}
\caption{The  plot of  the threshold  acceleration, $a_{\rm{max}}$,
is shown as a function of the distance between detectors and the
boundary for the three acceleration scenarios. Here, we have set
$\Omega\sigma=0.10$  and $\Delta{d}/\sigma=\{0.20\;,0.50\;,1.00\}$
in the left-to-right order.}\label{adethvsz}
\end{figure*}

To further understand the influence of the presence of the boundary on the harvesting-achievable range of
 the separation $\Delta{d}$  and the acceleration $a$ in the parameter space where entanglement harvesting is possible, we  define a new parameter $a_{\rm{max}}$
to denote the maximum  value of
the acceleration $a$, beyond which entanglement
harvesting does not occur any more, and plot  the dependence of $\Delta{d}_{\rm{max}}$ and $a_{\rm{max}}$ on the distance between the detectors and the boundary in Figs.~(\ref{ddethvsz}) and (\ref{adethvsz}) in all three acceleration scenarios.

As we can see from these figures,
$\Delta{d}_{\rm{max}}$ and $a_{\rm{max}}$ decrease with the
increasing distance between the detectors and the boundary. In particular,  the presence of the boundary could enhance  the harvesting-achievable range of $\Delta{d}$ for detectors in  acceleration or  at rest. This means that the positive impact of the boundary on the harvesting-achievable range is irrespective of the detectors' motion status.  However, the quantitative details of $\Delta{d}_{\rm{max}}$ differ slightly for different scenarios. As shown in Fig.~(\ref{ddethvsz}), the inertial case has a comparatively larger $\Delta{d}_{\rm{max}}$  than three acceleration scenarios for a certain fixed $\Delta{z}$, which can be simply attributed to the negative impact of the thermal noise that arises from  acceleration.
When $\Delta{z}\rightarrow\infty$, $\Delta{d}_{\rm{max}}$ and
$a_{\rm{max}}$ approach the corresponding values in the
free spacetime without boundaries. Therefore, we conclude that
the presence of the boundary enlarges the parameter space of $\Delta{d}$ and $a$ where the entanglement could be extracted successfully as compared to the case without a boundary, although it generally  degrades the  entanglement extracted by the accelerated detectors.

One can  observe from Fig.~(\ref{ddethvsz11}) that, for a small acceleration ($a\sigma\ll1$),
three solid curves approach the dashed curve, and this is expected since in
the limit of $a\rightarrow0$, we should recover the inertial case with detectors at rest. Differences also show up for different acceleration scenarios.
For a small acceleration,
one can see that the parallel acceleration case has a larger
$\Delta{d}_{\rm{max}}$ than both anti-parallel and
perpendicular acceleration cases, and the smallest  is  the anti-parallel
acceleration case. Physically, this means that the parallel acceleration provides
 spacious ``room" of separation for the detectors to  extract the entanglement successfully,
and the anti-parallel acceleration provides the least ``room" for entanglement extraction.  These
properties are however inconspicuous in the regime near the boundary. Note
that the dashed line, representing $\Delta{d}_{\rm{max}}/\sigma$ for two rest detectors, is always
above  the solid curves, which implies that acceleration always shortens the separation between the detectors that allows the entanglement harvesting to occur.

As acceleration grows,
 $\Delta{d}_{\rm{max}}$ will  decrease because
of the increased thermal noise caused by the Unruh effect. However,  the amount of the decrease differs for different acceleration scenarios so that when the acceleration is no longer small, the order of the magnitude of $\Delta{d}_{\rm{max}}$ changes (see Fig.~(\ref{ddethvsz22}) and Fig.~(\ref{ddethvsz33})). In place of the parallel acceleration, the
anti-parallel acceleration now becomes the optimum choice that has
the largest $\Delta{d}_{\rm{max}}$.
This can be easily understood in the same way as we did for the comprehension of  Figs.~(\ref{condzvsa}) and (\ref{condzvsd}).

In Fig.~(\ref{adethvsz}),  similar conclusion about
$a_{\rm{max}}$ can also be obtained. For a large $\Delta{d}/\sigma$,
the anti-parallel acceleration  has comparatively larger
$a_{\rm{max}}$ than both the parallel and perpendicular
acceleration, which means that the case of anti-parallel
acceleration has a larger range of acceleration for the detectors to harvest
entanglement in  the parameter space. However,  for a small
$\Delta{d}/\sigma$, the case of the parallel acceleration, especially in the  region far from the boundary, is the optimum choice that
has more ``room" to achieve the entanglement harvesting. Meanwhile, in the region close to
the boundary,  the values of $a_{\rm{max}}$ become considerably large, but the  differences of $a_{\rm{max}}$ among  three acceleration scenarios become tiny.

It is worth emphasizing that in the limit of $\Delta{z}\rightarrow0$
(i.e., location on the boundary plane), the corresponding
Wightman function Eq.~(\ref{wigh-1}) approaches  zero.  As
a result, the concurrence vanishes  and does not depend on the
parameters $\Delta{d}$ and $a$. Then the definition of
$\Delta{d_{\rm{max}}}$ and $\Delta{a_{\rm{max}}}$ would make no
physical sense. In the concrete  numerical
evaluation for Fig.~(\ref{ddethvsz})/(\ref{adethvsz}),   $\Delta{z}/\sigma<1/100$  is approximately  treated  as
two detectors being located on the boundary plane and no further numerical
calculation is  actually performed. Of course, if we attempt to
implement the numerical integration for $\Delta{z}/\sigma<1/100$, the vanishingly small integral  and the possible
oscillatory integrand with singularities should require extremely
high precision and very long computing time. Here, we choose not to
evaluate  $X$  and $P_D$ for
$\Delta{z}/\sigma<1/100$, since such an evaluation is not expected to qualitatively change our
conclusions. For clarity,  we summarize our main results  in Table.~(\ref{tab:addlabel1})and~(\ref{tab:addlabel2}) which follow.
\begin{table}[htbp]\footnotesize
  \centering
  \caption{Main results of the boundary influence for  accelerated and inertial rest detectors }
    \begin{tabular}{|p{21em}|p{19em}|p{17em}|}
    \hline
    \multicolumn{1}{|l|}{Harvested entanglement} & \multicolumn{1}{l|}{Three acceleration scenarios } & \multicolumn{1}{l|}{Rest case} \bigstrut\\
    \hline
    Inhibited by the boundary for $\Delta{z}\ll\sigma$  & Yes   & Yes \bigstrut\\
    \hline
     Peaks at $\Delta{z}\sim\sigma$ for a not large fixed $\Delta{d}$ & Yes, a comparatively smaller peak than the inertial case& Yes \bigstrut\\
    \hline
    Decreasing harvesting-achievable range of $\Delta{d}$ as $\Delta{z}$ increases & Yes   & Yes, a comparatively  larger harvesting-achievable range than the acceleration case  \bigstrut\\
    \hline
     Decreases as $\Delta{d}$ increases for a fixed $\Delta{z}$ & Yes   & Yes \bigstrut\\
    \hline
    \end{tabular} \label{tab:addlabel1}
\end{table}

\begin{table}[htbp]\footnotesize
  \centering
  \caption{Main results for three acceleration scenarios in comparison  }
    \begin{tabular}{|p{16.5em}|p{14em}|p{14em}|p{11.5em}|}
    \hline
    \multicolumn{1}{|l|}{Harvested entanglement} & \multicolumn{1}{l|}{Parallel acceleration} & \multicolumn{1}{l|}{Anti-parallel acceleration} & \multicolumn{1}{l|}{Perpendicular acceleration} \bigstrut\\
    \hline
    Degrades with increasing $\Delta{d}$ (acceleration $a$) for a large $a$ ($\Delta{d}$) at fixed $\Delta{z}$  & Most rapidly  & Rapidly & More rapidly  \bigstrut\\
    \hline
    The amount for a small (large) $a$ and $\Delta{d}$ at fixed $\Delta{z}$ & Most (least) amount & Least (most) amount & Medium amount   \bigstrut\\
    \hline
    The harvesting-achievable parameter space of $\Delta{d_{\rm{max}}}$ or $a_{\rm{max}}$  &  Largest (smallest) $\Delta{d_{\rm{max}}}$ or $a_{\rm{max}}$  for a fixed small (large) $a$ or $\Delta{d}$ &  Smallest (largest) $\Delta{d_{\rm{max}}}$ or $a_{\rm{max}}$ for a fixed small (large)  $a$ or $\Delta{d}$ & Medium $\Delta{d_{\rm{max}}}$ or $a_{\rm{max}}$  \bigstrut\\
    \hline
    \end{tabular} \label{tab:addlabel2}
\end{table}


\section{conclusion}
In this paper, we have explored  the phenomenon of entanglement
harvesting for  a pair of inertial as well as uniformly accelerated UDW detectors near a perfectly
reflecting boundary.
Three different acceleration scenarios, i.e., parallel, anti-parallel and mutually perpendicular
acceleration, are considered.  We find that the presence of the boundary
significantly influences on the entanglement harvesting of the accelerated
detectors. As a whole, the reflecting boundary inhibits
the  transitions of
detectors and nonlocal correlation of the fields  so that the harvested  entanglement would
degrade when two detectors are close to the boundary.   However, when the distance
between detectors and the boundary is comparable to  parameter
$\sigma$ which characterizes the interaction duration,  the harvested
entanglement may approach a peak, which
even goes beyond that without a boundary.
More interestingly, we find that the presence of the boundary, in all three
acceleration scenarios, could enlarge
the parameter space ( acceleration $a$ and detectors'
separation $\Delta{d}$ ) beyond which the entanglement harvesting
no longer occurs. These  conclusions also qualitatively hold for  inertial detectors at rest.

 A comparison of  three different acceleration scenarios reveals that the entanglement harvesting crucially depends on the distance between the detectors and the boundary,
the acceleration and the detectors' separation.  As far as the amount
of  the entanglement harvested is concerned, the detectors in parallel acceleration are likely to harvest the most entanglement in the case of a small
acceleration or small detectors' separation.  However, those in
anti-parallel acceleration  may harvest the most entanglement in the case of a large
acceleration and large detectors' separation.  For the parameter space that allows entanglement extraction to occur,
we find that, for a vanishing small acceleration, the harvesting-achievable
range of  detectors' separation for all three acceleration scenarios is basically the same  near
boundary.  However,  in the region far away from
the boundary,  the parallel acceleration  is the optimum choice that has the largest detectors' separation to
achieve  entanglement harvesting.  While for a not too small
acceleration, the anti-parallel acceleration becomes the optimum
choice that has the largest separation.  Similar
conclusions can be obtained for the harvesting-achievable range of
acceleration when entanglement harvesting is still possible with a certain fixed detectors' separation. Also, our direct numerical calculation instead of the saddle approximation
seems to indicate that acceleration always shortens the detectors' separation  that allows the entanglement harvesting to occur.

Finally, we have demonstrated how the presence of a boundary
affects the entanglement harvesting for two uniformly accelerated
detectors. We anticipate that  these methods can be used to
investigate the entanglement harvesting in other cases, such as two detectors with different  magnitudes of acceleration or with   acceleration along the  normal direction of
the boundary or even in curved background. It is quite a challenge to perform numerical evaluation
due to the complexity of the issue.  We would rather
leave such  studies  to future works.

\begin{acknowledgments}
 This work was supported in part by the NSFC under Grants No. 11690034 and No.12075084; and the Research Foundation of Education Bureau of Hunan Province, China under Grant No.20B371.
\end{acknowledgments}
\appendix
\section{Derivation of $P_D$ }\label{Derivation-PD}
To verify Eq.(\ref{PD-1}), let us begin from Eq.~(\ref{PAPB}).  Taking into consideration
the fact that the Wightman function~(\ref{wigh-1}) for
trajectory~(\ref{a-trj}) is only a function of the difference
between $\tau$ and $\tau'$ and letting $u=\tau$ and $s=\tau-\tau'$,
we have, after integrating $u$ firstly,
\begin{align}\label{PA-A1}
P_D&=\lambda^2\int_{-\infty}^{\infty}{du}\chi_D(u)\int_{-\infty}^{\infty}{ds}\chi_D(u-s)e^{-i\Omega{s}}W(s)\nonumber\\
&=\lambda^2\sqrt{\pi}\sigma\int_{-\infty}^{\infty}{ds}e^{-i\Omega{s}}e^{-s^2/(4\sigma^2)}W(s)\;.
\end{align}
By inserting Eq.~(\ref{a-trj}) into Eq.~(\ref{PA-A1}) and performing  some
simple algebraic manipulations,  the transition probability can be
written into two terms as
\begin{equation}
P_D=P_1+P_2\;,
\end{equation}
with the first term satisfying
\begin{align}\label{PA-A11}
P_1=-\frac{\lambda^2 a
\sigma}{8\pi^{3/2}}\int_{-\infty}^{\infty}d\tilde{s}\frac{{e}^{-i
\tilde{s}\beta}e^{-\tilde{s}^2\alpha}}{\sinh^2(\tilde{s}-i\epsilon)}\;,
\end{align}
and the second term
\begin{align}\label{PA-A21}
P_2=\frac{\lambda^2 a
\sigma}{8\pi^{3/2}}\int_{-\infty}^{\infty}d\tilde{s}\frac{{e}^{-i
\tilde{s}\beta}e^{-\tilde{s}^2\alpha}}{\sinh^2(\tilde{s}-i\epsilon
)-a^2\Delta z^2}\;,
\end{align}
where $\beta:=2\Omega/a $ and $\alpha:=1/(a\sigma)^2$.
 $P_1$ can be rewritten as
\begin{align}\label{PA-A12}
P_1&=-\frac{\lambda^2 a
\sigma}{8\pi^{3/2}}\int_{-\infty}^{\infty}d\tilde{s}\bigg[\frac{{e}^{-i
\tilde{s}\beta}e^{-\tilde{s}^2\alpha}}{\sinh^2(\tilde{s}-i\epsilon
)}-\frac{{e}^{-i
\tilde{s}\beta}e^{-\tilde{s}^2\alpha}}{(\tilde{s}-i\epsilon)^2}+\frac{{e}^{-i
\tilde{s}\beta}e^{-\tilde{s}^2\alpha}}{(\tilde{s}-i\epsilon)^2}\bigg] \nonumber\\
&=\frac{\lambda^2 a
\sigma}{4\pi^{3/2}}\int_{0}^{\infty}d\tilde{s}\frac{\cos(
\tilde{s}\beta)(\sinh^2\tilde{s}-\tilde{s}^2)}{\tilde{s}^2\sinh^2\tilde{s}}e^{-\tilde{s}^2\alpha}-\frac{\lambda^2
a \sigma}{8\pi^{3/2}}\int_{-\infty}^{\infty}d\tilde{s}\frac{{e}^{-i
\tilde{s}\beta}e^{-\tilde{s}^2\alpha}}{(\tilde{s}-i\epsilon)^2}\;.
\end{align}
 In second line of the above equation, we have neglected the factor $i\epsilon$ since the integral is now regular. While the
second term  can be re-expressed as
\begin{align}\label{PA-A13}
&-\frac{\lambda^2 a
\sigma}{8\pi^{3/2}}\int_{-\infty}^{\infty}d\tilde{s}\frac{{e}^{-i
\tilde{s}\beta}e^{-\tilde{s}^2\alpha}}{(\tilde{s}-i\epsilon )^2}
\nonumber\\
&=-\frac{\lambda^2 a
\sigma}{8\pi^{3/2}}\int_{-\infty}^{\infty}d\tilde{s}\frac{{e}^{-i
\tilde{s}\beta}e^{-\tilde{s}^2\alpha}}{\tilde{s}^2}+\frac{i\lambda^2
a \sigma}{8\pi^{1/2}}\int_{-\infty}^{\infty}d\tilde{s}{e}^{-i
\tilde{s}\beta}e^{-\tilde{s}^2\alpha}\delta^{(1)}(\tilde{s})\;.
\end{align}
Here, we have used the following identity that arises from the successive
differentiation  of  the Sokhotski formula,
 \begin{equation}\label{id0}
 \frac{1}{(x\pm{i}\epsilon)^n}=\frac{1}{x^n}\pm\frac{(-1)^n}{(n-1)!}{i\pi}\delta^{(n-1)}(x).
 \end{equation}
In addition, recalling the definition of a distribution $g$ acting on a
test function $f$
 \begin{equation}
\langle{g},{f}\rangle:=\int_{-\infty}^{\infty}g(x)f(x)dx\;,
\end{equation}
we have  the following identities for a distribution
function~\cite{EDU:2016-1,Bogolubov:1990}
\begin{equation}\label{id1}
\big\langle{\frac{1}{x}},{f(x)}\big\rangle={\rm{PV}}\int_{-\infty}^{\infty}\frac{f(x)}{x}dx\;,
\end{equation}
\begin{equation}\label{id2}
\big\langle{\frac{1}{x^2}},{f(x)}\big\rangle=\int_{0}^{\infty}dx\frac{f(x)+f(-x)-2f(0)}{x^2}\;,
\end{equation}
and
\begin{equation}\label{id3}
 \big\langle{\delta^{(n)}(x)},{f(x)}\big\rangle=(-1)^nf^{(n)}(0)\;,
 \end{equation}
 where $\rm{PV}$ denotes the principle value of an integral.
Thus, by using Eq.~(\ref{id2}) and Eq.~(\ref{id3}),  Eq.~(\ref{PA-A13}) can be further
 written in a simple form as
 \begin{align}\label{PA-A14}
&-\frac{\lambda^2 a
\sigma}{8\pi^{3/2}}\int_{-\infty}^{\infty}d\tilde{s}\frac{{e}^{-i
\tilde{s}\beta}e^{-\tilde{s}^2\alpha}}{(\tilde{s}-i\epsilon )^2}=\frac{\lambda^2}{4\pi}\Big[e^{-\Omega^2\sigma^2}-\sqrt{\pi}\Omega\sigma\;\rm{Erfc}\big(\Omega\sigma\big)\Big]\;.
\end{align}
As for $P_2$, it can be written as
\begin{align}\label{PA-A22}
P_2&=\frac{\lambda^2 a
\sigma}{8\pi^{3/2}}\int_{-\infty}^{0}d\tilde{s}\frac{{e}^{-i
\tilde{s}\beta}e^{-\tilde{s}^2\alpha}}{\sinh^2\tilde{s} -a^2\Delta
z^2+i\epsilon}+\frac{\lambda^2 a
\sigma}{8\pi^{3/2}}\int_{0}^{\infty}d\tilde{s}\frac{{e}^{-i
\tilde{s}\beta}e^{-\tilde{s}^2\alpha}}{\sinh^2\tilde{s} -a^2\Delta
z^2-i\epsilon}\nonumber\\&=\frac{\lambda^2 a
\sigma}{4\pi^{3/2}}{\rm{PV}}\int_{0}^{\infty}d\tilde{s}\frac{\cos(
\tilde{s}\beta)e^{-\tilde{s}^2\alpha}}{\sinh^2\tilde{s}-a^2\Delta
z^2}+\frac{\lambda^{2} a \sigma}{4
\sqrt{\pi}}\frac{e^{-\alpha{\tilde{s}^2}}\sin(\beta{\tilde{s}})}{\sinh(2\tilde{s})}\Bigg|_{\tilde{s}=\arcsinh(a\Delta{z})}\;,
\end{align}
where Eq.~(\ref{id0}) and  Eq.~(\ref{id3}) have been considered in
the last step. Combining Eqs.~(\ref{PA-A12}), (\ref{PA-A14}) and
(\ref{PA-A22}), one can easily verify the expression of the transition
probability given in Eq.~(\ref{PD-1}).

\def\ACP{AIP Conf. Proc.}
\def\AIHP{Ann. Inst. Henri. Poincar\'e}
\def\AJP{Amer. J. Phys.}
\def\AM{Ann. Math.}
\def\AP{Ann. Phys. (N.Y.)}
\def\APJ{Astrophys. J.}
\def\ASS{Astrophys. Space Sci.}
\def\ATMP{Adv. Theor. Math, Phys.}
\def\CJP{Can. J. Phys.}
\def\CMP{Commun. Math. Phys.}
\def\CPB{Chin. Phys. B}
\def\CPC{Chin. Phys. C}
\def\CPL{Chin. Phys. Lett.}
\def\CQG{Classcal Quantum Gravity}
\def\CTP{Commun. Theor. Phys.}
\def\EASPS{EAS Publ. Ser.}
\def\EPJC{Eur. Phys.  J. C.}
\def\EPL{Europhys. Lett.}
\def\GRG{Gen. Relativ. Gravit.}
\def\IJGMMP{Int. J. Geom. Methods Mod. Phys.}
\def\IJMPA{Int. J. Mod. Phys. A}
\def\IJMPD{Int. J. Mod. Phys. D}
\def\IJTP{Int. J. Theor. Phys.}
\def\JCAP{J. Cosmol. Astropart. Phys.}
\def\JGP{J. Geom. Phys.}
\def\JETP{J. Exp. Theor. Phys.}
\def\JHEP{J. High Energy Phys.}
\def\JMP{J. Math. Phys. (N.Y.)}
\def\JPA{J. Phys. A}
\def\JPCS{J. Phys. Conf. Ser.}
\def\JPSJ{J. Phys. Soc. Jap.}
\def\LMP{Lett. Math. Phys.}
\def\LNC{Lett. Nuovo Cim.}
\def\MPLA{Mod. Phys. Lett. A}
\def\NPB{Nucl. Phys. B}
\def\PCAM{Proc. Symp. Appl. Math.}
\def\PCPS{Proc. Cambridge Philos. Soc.}
\def\PDU{Phys. Dark Univ.}
\def\PLA{Phys. Lett. A}
\def\PLB{Phys. Lett. B}
\def\PR{Phys. Rev.}
\def\PRA{Phys. Rev. A}
\def\PRD{Phys. Rev. D}
\def\PRE{Phys. Rev. E}
\def\PRL{Phys. Rev. Lett.}
\def\PRX{Phys. Rev. X}
\def\PRSLA{Proc. Roy. Soc. Lond. A}
\def\PTP{Prog. Theor. Phys.}
\def\PRp{Phys. Rept.}
\def\RMP{Rev. Mod. Phys.}
\def\SB{Sci. Bull.}
\def\SPP{Springer Proc. Phys.}
\def\SRTU{Sci. Rep. Tohoku Univ.}
\def\ZPC{Zeit. Phys. Chem.}

\end{document}